# The Profiles of H$\beta$ and [O III] $\lambda$5007 in Radio-loud Quasars

M. S. Brotherton

*McDonald Observartory and Astronomy Department, University of Texas, Austin, TX 78712*
*E-mail: msb@astro.as.utexas.edu*

## ABSTRACT

Moderate-resolution (150-300 km s$^{-1}$) spectra of the H$\beta$–[O III] $\lambda$5007 region have been obtained for 41 radio-loud quasars in order to investigate relationships among their broad- and narrow-line profiles, radio structures, and X-ray properties. Comparable spectra from the literature have been included to form a data set of 60 radio-loud quasars with similar numbers of core- and lobe-dominated quasars.

A variety of statistical analyses have been performed, including rank-correlation tests and principal component analysis (PCA), identifying several strong, related trends. The dominant source of variation in the H$\beta$ line profile parameters is related to radio core dominance, which measures the orientation of the central engine in the relativistic beaming model (both the traditional R and a new core dominance parameter, R$_V$, are used). As core dominance increases, the equivalent widths EW$_{H\beta}$ and EW$_{5007}$ decrease, H$\beta$ wings (the red in particular) weaken and H$\beta$ becomes more symmetric, and Fe II/[OĨII] $\lambda$5007 increases. The above changes in H$\beta$ are interpreted as the spectrum-to-spectrum variation of a redshifted, very-broad component (FWHM $\sim$8000-9000 km s$^{-1}$). Secondary variations in H$\beta$ appear to involve a narrower component (FWHM $\sim$2000 km s$^{-1}$), with a strength unrelated to the core dominance or luminosity, but strongly correlated with the strength of [O III] $\lambda$5007. Another set of significantly correlated parameters involves the [O III] $\lambda$5007 profile, continuum luminosity, and spectral shape: [O III] $\lambda$5007 is broader when the optical-to-X-ray spectral index ($\alpha_{ox}$) and the optical luminosity are larger; the narrow-line (NLR) component of H$\beta$ also gets stronger relative to [O III] $\lambda$5007, and there is a component of variation in EW$_{H\beta}$ and EW$_{5007}$ in the sense of a Baldwin effect.

These results, as well as some noteworthy non-correlations, are briefly discussed in terms of underlying physical causes and other results from the literature. A quantitative comparison with other classes of AGNs is deferred to a forthcoming paper.

*Subject headings:* line: profiles, quasars: emission lines, quasars: general, radio sources: general

## 1. Introduction

The highly-collimated radio jets and extreme luminosities of quasars suggest a central engine with large angular momentum and a gravitational energy source: a massive black hole fed by an accretion disk (e.g., Lynden-Bell 1969). Spectropolarimetry and infrared spectroscopy revealing otherwise hidden broad lines in Seyfert 2 galaxies led to the development of axisymmetric "dusty torus" models (see Antonucci 1993 for a review). The single and double cones of extended [O III] $\lambda 5007$ emission and scattered light observed in AGNs also point to axisymmetry (e.g., Pogge 1989; Penston et al. 1990; Storchi-Bergmann & Bonatto 1991; Tadhunter et al. 1992; Morganti et al. 1992; Macchetto et al. 1994).

The jets of radio-loud quasars suggest a way to measure the orientation of AGNs, which permits the investigation of the geometry of the central engine and its environment. The relativistic beaming model for radio sources (e.g., Blandford & Rees 1979; Orr & Browne 1982; Hough & Readhead 1989) unifies core-dominated and lobe-dominated radio sources by means of orientation: core-dominant objects are those viewed close to the jet axis, while lobe-dominant objects are those viewed at larger angles. Doppler boosting of radio core emission suggests that the ratio of core-to-lobe radio flux density, R, can be used to measure the orientation of the jet to the line of sight. Thus correlations of optical continuum properties with R suggest axisymmetric continuum emission (e.g. Jackson et al. 1989; Impey & Tapia 1990; Wills et al. 1992a; Baker et al. 1994). Some emission-line properties (discussed below) also depend on core dominance – it is these relations that this investigation was intended to characterize.

A data set of even moderate size with both the signal-to-noise ratios and resolutions sufficient to measure the profiles of both the broad- and narrow-lines present in the spectra of low-redshift radio-loud quasars has not before been assembled. Many previously studied radio-loud samples have been deficient in core-dominant quasars, making correlations with core dominance suspect. The present data set was assembled to fill this void and answer key questions: How do broad and narrow profiles depend on core dominance? Are velocity fields asymmetric, or affected by the presence of jets? What trends are present among emission-line and continuum properties in radio-loud quasars, and how do these differ from radio-quiets? From lower luminosity AGNs? What physics govern the observed trends?

The rest of the introduction briefly summarizes some pertinent previous work. In § 2 the sample selection, observations, data reduction, and measurements are described, and comments are made for individual objects. In § 3 the results of various statistical tests are presented, including comparison of core- and lobe-dominant subsamples, correlation tests, and principal component analyses (PCAs). These are discussed briefly in § 4 in terms of underlying physical causes and other results in the literature. A quantitative comparison with other classes of AGNs is deferred to a later paper (Brotherton 1995). The conclusions are summarized in § 5.

### 1.1. Broad Lines

To understand quasars' powerful central engines we must examine their local environment: the broad-line region (BLR). The broad-emission lines in quasar spectra are believed to arise from the photoionization of dense ($\sim 10^{9-12}$ cm$^{-3}$), high-velocity ($\leq 10,000$ km s$^{-1}$) "clouds" within a parsec of the central continuum source. The cloud kinematics are determined by several factors of varying importance, including: the gravitational field of the central mass, radiation pressure from the ionizing continuum, the effects of drag from an inter-cloud medium, and the clouds' initial velocities. Differences in profiles of emission lines with differing ionization suggest complex emission regions, but how BLR clouds move, whether orbitally, radially, or chaotically, in a single continuous region or in multiple regions, is not yet completely determined. Pure radial motion, either inflow or outflow, appears unlikely (Wilkes & Carswell 1982; Kallman et al. 1992 – see also, e.g., Gaskell 1988, Koratkar & Gaskell 1991, Maoz et al. 1991, and Korista et al. 1995 for variability results in lower luminosity AGNs).

One of the strongest trends observed in quasar spectra is the inverse correlation between broad optical Fe II emission and narrow [O III] emission (e.g., Boroson & Green 1992). The effect has not yet been satisfactorily explained. Radio-loud quasars tend to have small Fe II/[O III] $\lambda 5007$ intensity ratios.

Balmer lines in lobe-dominated quasars are quite broad and have irregular profiles (e.g., Setti & Woltjer 1977; Miley & Miller 1979), including multiple peaks (Gaskell 1983). In fact, the line width of H$\beta$ (FWHM$_{H\beta}$) is inversely correlated with R in a man-ner attributable to BLR clouds having preferred velocities perpendicular to the radio jet direction (Wills & Browne 1986).

Some other differences in emission-line properties between core- and lobe-dominated quasars have been reported:

1. Core-dominated objects have larger Fe II/[O III] and H$\beta$/[O III] line intensity ratios (e.g., Steiner 1981), interpreted by Jackson & Browne (1991) in terms of axisymmetric broad-line emission.

2. Core-dominated objects have broader C III] $\lambda$1909 and C IV $\lambda$1549 lines in high-redshift objects (Corbin 1991). Wills, Fang, & Brotherton (1992b) find a marginally significant opposite result for C IV $\lambda$1549.

3. Core-dominated objects have C IV with excess emission in the blue wing compared with the red; lobe-dominated the reverse (Corbin 1991, a marginally significant result using radio spectral index rather than R to measure core dominance). However, there may also be luminosity dependencies; a significant correlation in the opposite sense is observed in a lower-luminosity sample (Wills et al. 1995a; see also Barthel, Tytler, & Thompson 1989).

4. The Balmer decrement (H$\gamma$/H$\beta$) may correlate with R (Jackson & Browne 1991), although Wills et al. (1995b) do not confirm this.

Observational differences between radio-loud and radio-quiet objects may indicate real differences in the nature of the emitting regions (perhaps the influence of the strong jet on the BLR), limiting the applicability of studies of one class to the other. Boroson (1992) argued against orientation effects in radio-quiet objects because the equivalent width of [O III] $\lambda$5007 (EW$_{5007}$) and FWHM$_{H\beta}$ appear uncorrelated – in radio-loud objects these quantities inversely correlate with R. This difference may result from radio-quiet objects' lack of an anisotropic (optical-synchrotron) continuum, not from intrinsic differences in BLR geometry. Some more direct differences include:

1. Radio-quiet quasars have H$\beta$ with either strong red or blue wings, but radio-loud objects usually show strong red wings (e.g., Sulentic 1989; Boroson & Green 1992).

2. Radio-loud quasars have weaker optical Fe II (e.g., Phillips 1977; Peterson, Foltz, & Byard 1981; Boroson & Green 1992).

3. Radio-loud quasars never have broad absorption lines (e.g., Morris et al. 1992).

4. Radio-loud quasars produce more X-rays relative to their optical luminosities (e.g., Browne & Murphy 1987; Wilkes et al. 1994).

5. Radio-loud quasars have narrower C IV $\lambda$1549 and C III] $\lambda$1909 emission lines with larger EWs, and smaller blueshifts relative to Mg II $\lambda$2798 (Steidel & Sargent 1991; Francis, Hooper, & Impey 1993; Brotherton et al. 1994a).

## 1.2. Narrow Lines

To understand the interaction between AGNs and their host galaxies we must examine the extended emission associated with the AGN phenomenon: the narrow-line region (NLR). The NLR exists on scales up to kiloparsecs with emission believed to result from photoionization of gas by the active nucleus (although shocks may play a role for AGNs of low luminosity). Compared with the BLR, NLR densities are lower ($<10^6$ cm$^{-3}$), as are velocities (hundreds of km s$^{-1}$).

Whittle, in a series of papers (1985a, b; 1992a, b, c), has investigated the kinematics of the NLR through profile studies of [O III] $\lambda$5007, attempting to unravel the roles of gravity, luminosity, and radio jets. In studying the correlations between W50 (W50 will always refer to the FWHM of [O III] $\lambda$5007) and the properties of both the host galaxy and the AGN, Whittle determined that the kinematics of the NLR in Seyfert galaxies are dominated by the galactic gravitational potential, but that there exists an additional acceleration mechanism in some radio-loud AGNs.

Busko & Steiner (1989) found that W50 correlated with the absolute visual nuclear magnitude in radio-quiet objects (both Seyferts and QSOs), but not radio-loud objects. They interpreted this as evidence for different line-broadening mechanisms in the NLRs of the two classes of objects.

Heckman, Miley, & Green (1984) studied the kinematics of the NLR using [O III] $\lambda$5007 profiles in a heterogeneous sample of mostly Seyfert galaxies. The kurtoses, asymmetries, and widths of [O III] $\lambda$5007 were similar in both quasars and active galaxies ($<W50> \approx 450$ km s$^{-1}$), implying no strong luminosity dependence for NLR kinematics. They reported a correlation between the widths of [O III] $\lambda 5007$ and the Balmer lines (see also Cohen 1983 and Whittle 1985b), and concluded that there must be a kinematic link between the two regions. Heckman et al. also claimed that strong blue wings on [O III] $\lambda 5007$ are rare among steep-spectrum (lobe-dominated) radio sources.

Jackson & Browne (1991) studied the [O III] $\lambda 5007$ profiles in 18 radio-loud quasars. They found a marginally significant correlation (at the 5% level) between asymmetry and core dominance consistent with the Heckman et al. result – only the core-dominant objects displayed a preference for strong blueward asymmetries, but they stressed that more observations of core-dominant quasars were required to confirm the result.

## 2. Observations and Measurements

### 2.1. The Data Set

The planned investigation required spectra of $\sim 50$ quasars, split between core- and lobe-dominant objects, for which H$\beta$ occurs at wavelengths less than 1$\mu$m. Quasars were selected from the Véron-Cetty & Véron catalog (1989) – a bias toward high-luminosity, broad-lined objects – according to the following criteria:

1. Radio-loud, with available R.
2. V $\leq$ 18.
3. z $\leq$ 0.95.
4. Declination $> -20°$.

A total of 41 spectra were observed at McDonald Observatory as described below. Previously published spectra of 19 additional quasars meeting the selection criteria were also gathered, to increase the size of the data set to 60 objects.

Table 1 gives the sample with additional data from the literature. Column (1) gives the 1950 coordinate names. Column (2) gives $M_{abs}$, the absolute magnitude based on V band measurements, from Véron-Cetty & Véron (1989). Column (3) gives the radio spectral index, S.I., between 6 and 11 cm, also from Véron-Cetty & Véron (1989). Column (4) gives log R, where R is the ratio of rest-frame radio core-to-lobe flux density at 5 GHz taken from Wills & Browne (1986), Wills et al. (1992), or Miller, Rawlings, & Saunders (1993). Column (5) gives a new parameter, log $R_V$, defined as the ratio of the radio core luminosity at 5 GHz rest frame to the optical luminosity (based on $M_{abs}$ in column 2). $R_V$ is an alternative measure of core dominance, employed because the extended radio luminosity is apprently affected by environmental factors unrelated to the central engine, resulting in a large scatter in R. $R_V$ has been defined: log $R_V$ = log($L_{core}/L_{opt}$) = log($L_{core}$) + $M_{abs}/2.5$ + 13.69. While $R_V$ does grossly distinguish between radio-loud and radio-quiet objects, it should not be thought of as a "radio-loudness" parameter, and it appears to measure the same property that R does. See Wills & Brotherton (1995) for more details. Column (6) lists the optical-to-X-ray (2 keV) spectral index ($\alpha_{ox}$: $F_\nu \propto \nu^{-\alpha_{ox}}$) from Wilkes et al. (1994) based on *Einstein* data. Columns (7) and (8) give the core and extended radio luminosities at 5 GHz rest frame. They have been calculated from R and the 6 cm fluxes from Véron-Cetty & Véron (1989). All luminosities and magnitudes assume $H_o = 50$ km s$^{-1}$ Mpc$^{-1}$ and $q_o = 0$.

The sample is incomplete, but large enough to obtain statistically significant results about trends present. The luminosity measurements (e.g., $M_{abs}$) are correlated with redshift – as expected for magnitude-limited samples – but the radio core dominance, R, is not significantly correlated with either redshift or optical luminosity. Since a radio source is brighter face-on (because of an optical-synchrotron component) than a physically-identical source with a larger inclination, the core-dominated quasars may be intrinsically fainter than their lobe-dominated counterparts. This may also be reflected by an inverse correlation between R and the extended radio luminosity (§ 3).

Spectra of radio-loud PG quasars from the Bright Quasar Sample (BQS, see Schmidt & Green 1983), a UV-excess, magnitude-limited sample, as observed by Boroson & Green (1992), are included in the data set. Note that a couple of these objects do not meet the magnitude cut-off typically used to define quasars (M < $-23$), but are included anyway. They used the KPNO 2.1m telescope and the Gold Spectrograph with a TI 800 × 800 CCD detector. The dispersion was 2.5 Å per pixel, and effective resolution was $\sim$ 7 Å using a 1.5″ slit.

Also included are high-resolution (2–3 Å) spectra from Jackson & Browne (1991). They observed a small sample of radio-loud quasars using the Intermediate Dispersion Spectrograph on the Isaac Newton Telescope. See Jackson & Browne (1991) and Jackson, Penston, & Perez (1991) for details.

## 2.2. New McDonald Observatory Data

Observations for 41 radio-loud quasars were obtained with the 2.7m telescope at McDonald Observatory and the Large Cassegrain Spectrograph. A thinned TI $800 \times 800$ CCD detector was used during February 1992 and earlier, and a thick Craf-Cassini $1024 \times 1024$ CCD detector was used during later runs. For most observations, a 600 l/mm grating gave a dispersion less than 2 Å per pixel with both detectors, but for the quasars with higher redshift ($z \sim 0.9$) a 300 l/mm grating was used. Because the primary goal was to study profiles and not spectrophotometry, a narrow slit (2″ EW) was used, giving a resolution of $\leq 5$ Å (the FWHM of the comparison lamp lines for the 600 l/mm gratings), and $\leq 10$ Å resolution for the 300 l/mm grating. Since the quasar image is centrally concentrated in the slit, the actual resolution is less. These set-ups resulted in resolutions of $\sim 150$ to 300 km s$^{-1}$ for the wavelengths observed, sufficient to resolve NLR lines.

Table 2 presents the observing log. Column (1) gives the 1950 coordinate names. Columns (2) and (3) give the redshifts and V magnitudes from Véron-Cetty & Véron (1989). Columns (3)–(6) give the UT dates, the gratings and slit widths used, and the exposure times for the observations.

## 2.3. Data Reduction and Measurements

The McDonald spectra were reduced using standard techniques with the IRAF NOAO package, including bias subtraction, flat-fielding, optimal extraction, wavelength and flux calibrations, and extinction correction. Multiple spectra were shifted in wavelength prior to combination when necessary. Hot sdO stars (typically from Massey & Gronwall 1990) were observed at similar airmasses to the quasars and used to divide out atmospheric features.

The broad H$\beta$ profiles were prepared for measurement by removing the narrow lines and the Fe II blends. The Fe II was removed using an empirical model based on the spectrum of I Zw 1 (PG 0050+124), a low-luminosity QSO with "narrow" broad lines and strong Fe II emission (the model of Boroson & Green 1992 was used). This model was broadened by convolution with Gaussian profiles of constant velocity width and scaled to fit the broad Fe II features at $\lambda\lambda 4450$-4700 and $\lambda\lambda 5150$-5350. The subtraction of this model removes Fe II from the H$\beta$ and [O III] $\lambda 5007$ profiles. As noted by Boroson & Green (1992), and seen here for some objects, the I Zw I Fe II multiplet ratios do not match all spectra; problems with Fe II subtraction are noted in the comments on individual objects (§ 2.4). Because the intention is to remove contaminating features, not to make exact Fe II measurements, no template has been subtracted when Fe II emission appears very weak.

The [O III] $\lambda 5007$ profile was used as a template to remove NLR H$\beta$ (hereafter H$\beta_n$) from the broad profile. The redshift of [O III] $\lambda 5007$ was used to determine the wavelength of H$\beta_n$. Despite good instrumental resolution, the transition between broad and narrow component is not always perceptible. In these cases an upper limit to the strength of H$\beta_n$ was estimated by finding the part of the H$\beta$ peak that reproduced the [O III] FWHM, and this was used to scale the narrow [O III] $\lambda 5007$ profile to remove H$\beta_n$. The values of [O III] $\lambda 5007$/H$\beta_n$ will be given statistically and included in the correlation analysis. The mean line intensity ratio of [O III] $\lambda 5007$ to H$\beta_n$ was 11.0±0.7, consistent with previous observations (e.g., Osterbrock 1989).

Figures 1a through 1g show the reduced spectra on both observed and rest frame wavelength scales, rebinned to approximately two pixels per resolution element. The rest frame wavelength scale is based on the redshift determined by fitting a Gaussian to the upper third of [O III] $\lambda 5007$.

### 2.3.1. Broad-Line Measurements

Measurements of broad H$\beta$ are given in Table 3. Column (1) gives the object's name. Column (2) gives the rest frame EW$_{H\beta}$ in Å. Columns (3)–(5) give the line width of H$\beta$ in km s$^{-1}$ at the 1/4, 1/2, and 3/4 levels, respectively. Column (6) gives the shift of H$\beta$, defined as the central wavelength at the 3/4 level relative to the 4861.3 Å in the rest frame ($\lambda_c(3/4)$) normalized by FWHM$_{H\beta}$. Redshifts are positive, blueshifts negative. Column (7) gives Boroson & Green's shape parameter defined as Shape = (FW1/4M + FW3/4M)/2×FWHM. Column (8) gives the more sensitive shape parameter of Wills et al. (1993) defined as Shape2 = (FW1/4M − FW3/4M)/FWHM. Larger Shape2 parameters indicate more sharply-peaked lines. Column (9) is the asymmetry parameter as defined by De Robertis (1985): $\text{Asymm}_{H\beta} = (\lambda_c(3/4) - \lambda_c(1/4))/\text{FWHM}_{H\beta}$. Negative values indicate stronger red wings. Care is sometimes needed since $\text{Asymm}_{H\beta}$ is not relative to the $H\beta_n$ peak: $H\beta$ profiles with excess flux longward of 4861.3Å rest frame and redshifted peaks can have an $\text{Asymm}_{H\beta}$ characteristic of a strong blue wing (e.g., 0710+118). Column (10) gives $\text{EW}_{5007}$, column (11) $\text{EW}_{FeII}$ as measured from the unbroadened template between 4434Å and 4684Å; all EWs are given in the rest frame and referenced to the Fe II-subtracted continuum level at 4861.3Å. The above parameters were chosen to facilitate comparisons with Boroson & Green (1992).

Column (12) gives a flag, A, B, C, or D, to indicate the quality of the spectra and the associated measurements. The "A" spectra (continuum signal-to-noise ratio > 20 per pixel) have errors of approximately 5% for simple line measurements like $\text{EW}_{H\beta}$ and $\text{FWHM}_{H\beta}$, with higher uncertainties for compound measurements such as $\text{Asymm}_{H\beta}$. The primary source of error is in the continuum placement (a locally-determined power law has been assumed). The "B" and "C" spectra have errors on the order of 5–15% and signal-to-noise ratios less than 20 per pixel in the continuum. The "D" spectra are generally unsuitable for all but estimates of EW and FWHM.

He II $\lambda 4686$ is clearly present in only about 1/3 of new spectra, and its strength is not given, although the presence of broad and narrow He II in individual objects is noted in column (13) of Table 3. Boroson & Green (1992) found an inverse correlation between the line intensity ratio He II $\lambda 4686/H\beta$ and luminosity, consistent with the weakness of He II $\lambda 4686$ in the more luminous objects in this sample.

Several spectra show a red "shelf" extending from $H\beta$ under [O III] $\lambda\lambda 4959, 5007$. It is assumed that this emission is from $H\beta$, but the possibility that this is broad [O III] should be mentioned (Meyers & Peterson 1985; van Groningen & de Bruyn 1989). The suspect objects are noted in § 2.4.

*2.3.2. Narrow [O III] $\lambda 5007$ Measurements*

Profile measurements of [O III] $\lambda 5007$ were made for the McDonald and Jackson & Browne spectra, and are given in Table 4. Column (1) gives the object name. Column (2) repeats $\text{EW}_{5007}$ from Table 3. Column (3) gives a measure of the $\lambda 5007$ absolute magnitude defined as $M_{OIII} = M_{abs} - 2.5 \log(\text{EW}_{5007})$.

Column (4) gives the line intensity ratio [O III] $\lambda 5007/H\beta$. Column (5) gives the FWHM of [O III] $\lambda 5007$, W50, in km s$^{-1}$. Column (6) gives the full width at 20% of the peak flux density, W20, in km s$^{-1}$. Column (7) gives the ratio of W50 to W20. Both W50 and W20 have been corrected for resolution effects by subtracting in quadrature the widths of the comparison lines at 50% and 20%, respectively. Column (8) gives AI20, a measure of the asymmetry of [O III] $\lambda 5007$ defined as the difference between the half widths of the blue and red wings at the 20% level, normalized by W50. A straight-forward correction for resolution effects is not available for AI20. Columns (9) and (10) give area parameters: IPV20, the inter-percentile velocity at 20%, is the width in km s$^{-1}$ between the wavelengths for which 10% of the total line intensity is present in each wing; Asy20 is the average of those wavelengths normalized by W50. These have been corrected for resolution effects according to the prescription developed by Whittle (1985b). Column (11) gives the FWHM of the estimated instrumental resolution assuming a uniformly illuminated aperture. This is a slight overestimate of the true resolution, but the lines are essentially resolved and the corrections are small, e.g., $\text{FWHM}_i$ and W50 are not significantly correlated.

### 2.4. Comments on Individual Objects

0044+030: The spectrum of Grandi (1982) has been used to estimate the Fe II template to subtract as a low-resolution spectrum was not otherwise available.

0115+027: The dip near 7600 Å is from incomplete atmospheric correction.

0134+329: From Jackson & Browne (1991), a compact steep-spectrum (CSS) radio source. Measurements were made of the spectrum obtained with the slit at a position angle of 344°, parallel to the inner radio axis.

0710+118: Has the broadest profile and the smallest R in the sample. The other lines in this object's spectrum are also quite broad (Wills et. al 1995a).

0736+017: The removal of Fe II blended with the [O III] $\lambda\lambda 4959, 5007$ profiles was poor, making area parameter measurements uncertain.

0742+318: Has a very broad $H\beta$ profile, probably double-peaked.

0837−120: Broad He II $\lambda 4686$ is blended with the blue wing of $H\beta$, and a single Gaussian fit was sub-tracted prior to measuring H$\beta$.

0838+133: Atmospheric A band was not perfectly corrected.

0903+169: Atmospheric B band absorption occurs just to the red of the H$\beta$ peak, but it divided out well. Other lines in 0903+169's spectrum show similar structure on the red wing (Wills et al. 1995a).

0955+326: Has very strong and "broad" narrow lines, making it difficult to estimate H$\beta_n$. Also, the Fe II multiplet ratios differ from the template, although the Fe II emission appears weak. After removal of H$\beta_n$, FWHM$_{H\beta}$ = 1400 km s$^{-1}$ is extreme (although the uncertainties are large), significantly narrower than the 4800 km s$^{-1}$ of Mg II $\lambda$2798 reported by Wills & Browne (1986). In Figure 2 spectra of 0955+326 from two epochs are plotted, scaled such that both have the same flux in [O III] $\lambda$5007. Broad H$\beta$ appears narrower than it was in the past.

1020−103: This is another object with a very broad H$\beta$ line, also having a double-peaked profile (as previously noted by Stirpe 1990). Like 0742+318, it is also core-dominant.

1028+313: The H$\beta$ profile has a broad, shelf-like red wing which could possibly be broad [O III] emission.

1103−006: This object was also observed by Boroson & Green (1992), but a better spectrum was obtained from McDonald Observatory.

1150+497: Another object with a red "shelf" on H$\beta$.

1156+295: The object was much fainter than the V = 14.4 given by Véron-Cetty & Véron (1989). The V-based M$_{abs}$ is not used in the statistical analyses.

1250+568: Another object with a red shelf on H$\beta$.

1340+290: Radio structure was not found in the literature. The [O III] $\lambda$5007/H$\beta_n$ is large ($\sim$ 30).

1510−089: This object is among the strongest Fe II emitters in the sample. The template does not fit the multiplet ratios perfectly, but it does remove much of the gross structure.

1618+177: Atmospheric A band absorption occurs just to the red of the H$\beta$ peak, and did not divide out perfectly.

1704+608: [O III] profile measurements were made from the spectrum of Jackson & Browne (1991) and are based on $\lambda$4959 instead of $\lambda$5007 because of uncorrected atmospheric absorption.

## 3. Analysis and Results

### 3.1. Mean Properties and Distributions

Figure 3 gives histograms of all quantities for both the core- (dotted line – log R > 0; filled circles in later plots) and lobe-dominated (solid line – log R < 0; open circles in later plots) subsamples. Table 5 presents the mean properties of the whole sample and the core-dominant and lobe-dominant subsamples. The H$\beta$ profiles tend to have strong red wings, small peak shifts (slightly to the red), and sharper peaks and broader wings than Gaussians (compare <Shape2$_{H\beta}$> = 1.16 with 0.77 for a Gaussian and 1.15 for a Lorentzian). The H$\beta$ profile asymmetry increases at lower intensity levels. The [O III] $\lambda$5007 profiles tend to have strong blue wings.

Also given in Table 5 are the results of Wilcoxon rank sum tests (also known as the Mann-Whitney test) that were used to determine whether the medians of the core-dominant distributions differ from those of the lobe-dominant subsample. As expected, the medians of the core dominance measures, R, R$_V$, S.I., and L$_{core}$, differ quite strongly, but L$_{ext}$, believed to be an isotropic parameter, differs at the 3 $\sigma$ level. This is probably a selection bias as discussed in § 2. A few other marginally significant differences are present: compared with the core-dominant objects, the lobe-dominant objects have larger $\alpha_{ox}$, larger H$\beta$ line widths, larger EW$_{5007}$, and weaker optical Fe II. These differences have been reported previously, and will be discussed in § 4.

### 3.2. Correlation Analysis

Table 6 gives the results of rank-correlation tests involving the tabulated parameters. The probability of a false correlation arising by chance between two parameters is calculated using the Spearman rank correlation coefficient and Student's t statistic. When this probability less than 10%, it is given beneath the correlation coefficient. With this many tests (378), several statistically significant correlations are expected to be spurious. The actual number of independent tests is much smaller ($\sim$100) since many of the parameters measure similar properties (e.g., S.I., R, and R$_V$; W20 and IPV20; Shape and Shape2; AI20 and Asy20). Discounting redundant correlations leaves about 30 "real" correlations, at the 1% level or better, that may be of physical interest, only one of which is expected by chance.

### 3.2.1. Correlations with Core Dominance

Significant correlations are present, confirming several previous studies. As reported by Wills & Browne (1986), FWHM$_{H\beta}$ inversely correlates with R. A refinement is that the correlation is weaker for higher intensity levels – FW1/4M$_{H\beta}$ is also inversely correlated with R, but FW3/4M$_{H\beta}$ is not (Fig. 4), although increased scatter from measuring errors could reduce the significance of the correlation for FW3/4M$_{H\beta}$. Also plotted are the correlations with R$_V$ which are *much* more significant ($\sim 6\,\sigma$ compared with $\sim 3\,\sigma$ – see Table 6). H$\beta$ line widths and M$_{abs}$ are uncorrelated, so the strength of the correlations with R$_V$ are not artificially induced. Another new significant correlation is found between Shape2$_{H\beta}$ and R$_V$, indicating real differences in the relations between core dominance and line widths at different levels.

The H$\beta$ line widths have larger correlation coefficients when taken with L$_{core}$ than with R, and are uncorrelated with the presumably isotropic L$_{ext}$. Both R$_V$ and L$_{core}$ correlate with H$\beta$ line width at similar significance levels. Because R$_V$ is not correlated with L$_{ext}$ and R is, the most likely interpretation seems to be that scatter in L$_{ext}$ (possibly from environment factors such as "hot spots," as described by Bridle et al. 1994) weakens the correlation between R and H$\beta$ line width. More discussion of the R$_V$ parameter is given by Wills & Brotherton (1995).

Other lines also have a dependence on core dominance. There exists a marginal (3%) inverse correlation between EW$_{5007}$ and R (see also, e.g., Jackson et al. 1989; Baker et al. 1994). This is shown in Figure 5, and appears consistent with the idea that anisotropic continuum beaming in the core-dominant objects alone causes the correlation (e.g., the models of Browne & Murphy 1987 and the corrections of Wills et al. 1992a). Both EW$_{FeII}$ and Fe II/H$\beta$ are correlated with R, as found previously (e.g., Joly 1991; Jackson & Browne 1991).

The [O III] profile parameters do not correlate well with the core dominance. Dusty NLR models invoking radial flow predict increasing profile asymmetries with R. Jackson & Browne (1991) found a marginal correlation between AI20 and R for a sample deficient in core-dominated objects. In this larger sample, the correlation is not significant.

### 3.2.2. Correlations among H$\beta$ Profile Parameters

There is a strong correlation between FWHM$_{H\beta}$ and EW$_{H\beta}$ (Fig. 6) noted previously for both low-luminosity AGNs (Gaskell 1985) and high-luminosity AGNs (Wills et al. 1992b). This is striking because the same quantities are inversely correlated for C IV $\lambda$1549 (Wills et al. 1993) and Ly $\alpha$ (Francis et al. 1992). There are differences in this relation when radio-quiet and radio-loud objects are considered separately (Brotherton 1995) which weaken the correlation in data sets that include both types of objects (the reason that Boroson & Green underestimated the strength of the correlation). Two other significant correlations between EW$_{H\beta}$ and [O III] $\lambda$5007 profile measurements are also shown in Figure 6.

Asymm$_{H\beta}$ is correlated with Shape2$_{H\beta}$: the lines with the broadest wings tend to have stronger red wings; Shape2 correlates with FW1/4M, but not significantly with FWHM or FW3/4M. The plots of the correlations between Asymm$_{H\beta}$, FW1/4M$_{H\beta}$, and Shape2$_{H\beta}$ are shown in Figure 7. Boroson & Green (1992) previously noted that the degree of H$\beta$ asymmetry is larger at lower intensity levels.

The strong inverse correlation between the Fe II and [O III] $\lambda$5007 strengths (e.g., Boroson & Green 1992) is also present. Also confirmed is that the H$\beta$ line asymmetry is weakly tied to the Fe II – [O III] inverse correlation: at small Fe II/[O III] $\lambda$5007, primarily red H$\beta$ asymmetries and symmetric [O III] $\lambda$5007 profiles are seen; at large Fe II/[O III] $\lambda$5007, [O III] $\lambda$5007 profiles have stronger blue wings. If Fe II has not been sufficiently corrected for, contamination could lead to [O III] profile asymmetries.

### 3.2.3. Correlations with the [O III] $\lambda$5007 Profile

There are a number of strong correlations involving [O III] $\lambda$5007 line parameters (widths at 20%, asymmetry, and luminosity) and $\alpha_{ox}$ (Fig. 8). The [O III] $\lambda$5007 line width is also correlated with M$_{abs}$ (Fig.9). Busko & Steiner (1989), studying a slightly less luminous range of AGNs, found this correlation to hold for radio-quiet objects but not for radio-loud objects. The presence of the correlation is not surprising given that there exists a non-linear relationship between the X-ray and optical luminosity (e.g., Avni & Tannanbaum 1982; Kriss & Canizares 1985; Browne & Murphy 1987) that manifests in the correlation between $\alpha_{ox}$ and M$_{abs}$. There is a marginal correlation between the H$\beta$ and [O III] $\lambda$5007 asym-metries, even though stronger red wings are typical for H$\beta$ and stronger blue wings are typical for [O III] $\lambda$5007.

The line intensity ratio [O III] $\lambda$5007/H$\beta_n$ is significantly correlated with $\alpha_{ox}$ (Fig. 10), and also with W20. These latter correlations could be the result of the method used to estimate the strength of H$\beta_n$, although in this case it would be strange that W50 is not also correlated with $\lambda$5007/H$\beta_n$.

All measurements of the [O III] $\lambda$5007 line width (W50, W20, and IPV20), are strongly correlated. Whittle (1992d) showed this for Seyferts and suggested a single acceleration mechanism for gas of both high and low velocity dispersion. The kurtosis measure, W50/W20, is significantly correlated with EW$_{H\beta}$ (Fig. 6, right panel), so there may be some influence from a second mechanism. The luminosity indicators are more strongly correlated with W50 than with W20.

Again, there are some important non-correlations to note. Jackson & Browne's (1991) investigation included a subsample of 18 quasars for which they obtained much higher resolution spectra (2-3Å). They found a marginal correlation (5%) between AI20 and R, but they had few objects with large R (See also Heckman et al. 1984). The current sample does not show this.

The [O III] $\lambda$5007 and H$\beta$ line widths are not correlated. Previously, Heckman et al. (1984) studied [O III] $\lambda$5007 profiles in a sample of 34 AGNs (both radio-loud and radio-quiet) with moderate-resolution spectra. Most of their objects were of lower luminosity than those here. They found a correlation between the line widths of [O III] and H$\alpha$ (see also Cohen 1983 and Whittle 1985b). Much of the correlation can be attributed to objects with W50 of less than $\sim$ 300 km s$^{-1}$ and FWHM$_{H\alpha}$ < 2000 km s$^{-1}$. Few Balmer lines in radio-loud quasars have FWHM < 2000 km s$^{-1}$ (Wills & Browne 1986). The correlation appears to break down for large line widths. This would imply that in high-luminosity objects, different line-broadening mechanisms are at work in the NLR and BLR.

### 3.3. Principal Component Analysis (PCA)

In an attempt to simplify the large number of correlations in terms of a smaller number of primary correlations, PCA has been employed. In essence, PCA defines eigenvectors composed of linear combinations of the input parameters; these eigenvectors are determined by finding the coordinate directions that account for the most variation in the input parameters, in order of their importance: the so-called principal components (the 1st, 2nd, etc.). Ideally, the original set of input parameters is reduced to a smaller number of new parameters (components), each representing the variation of a single underlying physical parameter.

Numerous PCA trials were performed using different sets of quasi-independent variables. Tables 7 and 8 present two illustrative cases. For both, rank values were used rather than the measured values in order to minimize the influence of extreme objects (e.g., 0710+118); these non-parametric analyses yield results that are qualitatively similar to the equivalent parametric analyses. A small subset of properties is used for the PCA in Table 7, allowing the inclusion of 53 objects – most of the sample. Table 8 presents the results of an analysis using more parameters, but complete data are available for only 30 objects (limited by the objects with high-resolution spectra and $\alpha_{ox}$ – there are similar numbers of core- and lobe-dominated objects). The important difference in Table 8 is the inclusion of the X-ray and [O III] profile data. The projections onto the input properties of the 6 eigenvectors accounting for the highest percentages of variation, and their cumulative percentage of variation, are given. The first few eigenvectors for each analysis are described below.

Eigenvector 1 in Table 7 involves most of the input variables except M$_{abs}$, Shift$_{H\beta}$, and $\lambda$5007/H$\beta$. The core dominance, R$_V$ in this analysis, suggests that the eigenvector may be related to orientation. Inversely correlated with R$_V$ are FWHM$_{H\beta}$, EW$_{H\beta}$, Shape2$_{H\beta}$, while positively correlated are Asymm$_{H\beta}$ and Fe II/[O III] $\lambda$5007. The variation of the H$\beta$ profile measures suggests the variation of a very broad, redshifted emission component that does not affect Shift$_{H\beta}$ but does affect FWHM$_{H\beta}$, Asymm$_{H\beta}$, and Shape$_{H\beta}$. Because EW$_{H\beta}$ has a significant projection on the vector but M$_{abs}$ does not, this argues for a changing line strength, not simply a redistribution of flux and a changing continuum. This broad, redshifted component decreases in strength with increasing core dominance.

Eigenvector 2 in Table 7 involves the H$\beta$ line width, shape, and peak shift as well as the strength of [O III] $\lambda$5007. This vector can be interpreted as arising from a narrow component of broad H$\beta$ (not the NLR, which has been removed) with a peak less redshifted relative to the remainder of the H$\beta$ emission and a relatively minor contribution to the line strength. The eigenvector is dominated by [O III] $\lambda$5007 emission, and has little dependence on either core dominance or luminosity.

Eigenvector 3 in Table 7 involves more changes in the H$\beta$ profile in addition to a strong inverse correlation between Fe II and [O III]. Eigenvector 4 is dominated by redshift and/or luminosity effects.

Eigenvector 1 in Table 8 is dominated by the correlations between $\alpha_{ox}$, M$_{abs}$, the EWs of H$\beta$ and [O III], and a measure of the [O III] line width, IPV20. More optically-luminous objects have weaker EWs (a Baldwin effect), correspondingly weaker X-ray luminosity, and broader narrow lines. Because the direct correlations between the EWs and M$_{abs}$ are weak (see Table 6), other variables might be obscuring a clear Baldwin effect. The intensity ratio [O III] $\lambda$5007/H$\beta_n$ is also correlated with this vector.

Since the PCA in Table 8 contains only half the sample, additional eigenvectors will not be discussed except to say that eigenvector 2 is similar to Table 7's eigenvector 1 insofar as it is strongly associated with R$_V$, EW$_{H\beta}$, and FWHM$_{H\beta}$. The equivalence of eigenvectors between the two tables is not completely clear, however, showing the limitations of PCA beyond the first couple of components for a data set of this kind.

## 4. Discussion

PCA provides a method of identifying related correlations that may correspond to independent, underlying physical parameters that determine the object-to-object variation of quasar properties. This discussion will focus on interpreting the main results of the PCA in terms of: emission of H$\beta$ from a very-broad-line region (VBLR$_{H\beta}$) related to core dominance (Table 7, eigenvector 1); emission of H$\beta$ from an intermediate-line region (ILR$_{H\beta}$) related to NLR emission (Table 7, eigenvector 2); and a luminosity/$\alpha_{ox}$ effect related to NLR kinematics (Table 8, eigenvector 1). The terminology of the ILR and VBLR is motivated both by the PCA and by the discovery of similar, but not necessarily related, emission components in broad UV lines (e.g., Francis et al. 1992; Wills et al. 1993; Brotherton et al. 1994a, b). It is important to see how radio-quiet objects fit into this characterization, whether radio-loudness is a fourth parameter, or whether radio-quiet objects represent an extreme of one or more of the above three effects. A detailed, quantitative comparison between this sample and radio-quiet objects from Boroson & Green (1992) will be reported elsewhere (Brotherton 1995).

### 4.1. Core Dominance and a Very-Broad-Line Region (VBLR$_{H\beta}$)

Eigenvector 1 of the PCA given in Table 7 is dominated by the core dominance (orientation in beaming models) and VBLR$_{H\beta}$. As core dominance decreases, the strength of emission from the VBLR$_{H\beta}$ increases, leading to larger EW$_{H\beta}$, FWHM$_{H\beta}$, and Shape2$_{H\beta}$ parameters, smaller Asymm$_{H\beta}$ (stronger red wings), while Fe II/[O III] decreases. Figure 11 shows the difference between the mean spectra formed from those objects with FWHM$_{H\beta}$ less than the mean (4700 km s$^{-1}$) and those greater than the mean. The broad dip has FWHM $\sim$ 8000-9000 km s$^{-1}$ and is consistent with, but not necessarily independent evidence for, the expected VBLR$_{H\beta}$ component. Figure 11 provides a way of looking at how the real spectra change with line width rather than trying to infer this from a few parameters.

Several explanations are possible for the variation of emission from a VBLR with core dominance, a simple one being that H$\beta$ is emitted anisotropically from the highest velocity clouds but isotropically from lower velocity clouds. Some elements needed in future models include clouds with a range of optical depths, anisotropic line and/or continuum emission, anisotropic velocity fields, or if the variation of the core dominance in this vector is not the result of beaming and orientation, the interaction between BLR clouds and jets, or simply intrinsic variation in core luminosities.

Others have put forth explanations for correlations associated with the core dominance and VBLR$_{H\beta}$. Wills & Browne (1986) suggested that the inverse correlation between R and FWHM$_{H\beta}$ was the result of a redistribution of line flux with viewing angle of a disk-shaped BLR, and EW$_{H\beta}$ varies with an axisymmetric continuum (see also the "Bipolar Model" of Wills et al. 1993). Because the projection of this eigenvector along M$_{abs}$ is so small, the variation in EW$_{H\beta}$ is primarily the result of a changing line strength, *not* a changing continuum, making the Wills & Browne explanation less likely than the VBLR$_{H\beta}$ explanation. Another idea was proposed by Gaskell (1985), who interpreted the correlation between FWHM$_{H\beta}$ and EW$_{H\beta}$ as evidence for radiative acceleration.

Variability has provided perhaps the best evidence for independent behavior in high-velocity gas emitting H$\beta$ from some type of VBLR – but not necessarily the same region suggested by the PCA or the analysis by Brotherton et al. (1994b) . Ferland, Korista, & Peterson (1990) studied the low-luminosity AGN Mrk 590 and noted that the wings of H$\beta$ maintained a constant luminosity while the continuum luminosity increased by 0.5 dex. The line core luminosity followed the continuum. Assuming that H$\beta$ arises from photoionization, they concluded that there was an optically-thin VBLR shrouding the central engine and distinct from a more distant optically-thick line-emitting region.

The VBLR$_{H\beta}$ component differs in a few key ways from the VBLR identified in the UV lines of radio-quiet quasars (Brotherton et al. 1994b). That VBLR was blueshifted and led to profiles with strong blue wings, not the red wings of the present sample. Barthel et al.'s (1989) radio-loud sample had many C IV $\lambda$1549 profiles with strong red wings, so it seems that asymmetry (and the redshift of the VBLR component) may depend on radio-loudness (see also Wills et al. 1995a). Sulentic (1989) and Boroson & Green (1992) have pointed out that radio-loud objects have H$\beta$ with strong red wings, but radio-quiet objects show profiles with either wing stronger.

Several objects in the present sample show He II $\lambda$4686 broader than H$\beta$ (see also Boroson & Green 1992, e.g., PG 0003+158). Marziani & Sulentic (1993) noted the greater width of He II $\lambda$4686 (compared with H$\beta$) in PG 1138+222 – using photoionization models they derived the characteristics of the high-velocity gas, which turned out to be similar to those of the VBLR of Brotherton et al. (1994b).

### 4.2. An Intermediate-Line Region (ILR$_{H\beta}$)

Eigenvector 2 in Table 7 is identified with an ILR$_{H\beta}$ component less redshifted relative to the remainder of the broad H$\beta$ emission (redshifted based on the mean Asymm$_{H\beta}$ and Shift$_{H\beta}$). Variations in a weak, narrow component can lead to significant changes in FWHM and the peak redshift without much changing the overall line strength. The difference spectrum in Figure 11 shows an ILR component, with a redshift of ~150 km s$^{-1}$ and a FWHM of 2000 km s$^{-1}$. The narrow [O III] lines are also seen to be stronger in the objects with narrower H$\beta$ profiles.

The properties of ILR$_{H\beta}$ are consistent with the ILR of Brotherton et al. (1994b) in terms of redshift, line width, and the lack of correlations with radio spectral index (another measure of the core dominance) or optical luminosity. Their ILR and VBLR photoionization models predict that for H$\beta$ the VBLR contributes more to the overall line strength than the ILR, which seems borne out. Investigation of the relationship of the components in H$\beta$ with the ILR and VBLR will require observations of lines of different excitation (e.g., both C IV $\lambda$1549 and H$\beta$) in the same spectra of both radio-loud and radio-quiet objects. Several such projects are underway (e.g., Wills et al. 1995a,b).

Some possible origins of an independently varying ILR component in UV lines were discussed by Francis et al. (1992) and Wills et al. (1993). These may also be applicable to ILR$_{H\beta}$.

### 4.3. Luminosity Effects

Eigenvector 1 in Table 8 involves luminosity effects: as the optical luminosity increases, the EWs of H$\beta$ and [O III] $\lambda$5007 decrease, while the [O III] line widths and $\alpha_{ox}$ increase (the X-ray luminosity does not increase as fast as the optical). The line intensity ratio $\lambda$5007/H$\beta_n$ is also correlated with this vector.

Steiner (1981) previously noted a weakening of [O III] $\lambda$5007 (relative to H$\beta$) with increasing luminosity for certain classes of AGNs. The clearest "[O III] Baldwin effect" was seen in objects with optical Fe II emission (primarily radio-quiet objects), which is consistent with the weakness of the effect in this radio-loud data set.

One simple interpretation is that as AGN luminosity increases, the ionizing continuum becomes softer – previously suggested on observational grounds. Certain line ratio correlations with luminosity (e.g., Mushotzky & Ferland 1984; Kinney et al. 1987) and the inverse correlation between the strength of He II $\lambda$4686 and luminosity (eigenvector 2 of the PCA of Boroson & Green 1992) can be explained by an inverse correlation between the ionization parameter and luminosity. This result may also extend to Seyfert galaxies (Ho, Shields, & Filippenko 1993).

Netzer, Laor, & Gondhalekar (1991) proposed an explanation for the Baldwin effect based on the inverse relation between radiation hardness and luminosity in accretion disks. A more luminous, more massive disk emits more ionizing photons, and if the NLR is radiation-bounded, then the NLR will become larger as more low-velocity gas is ionized, resulting in stronger [O III] line cores relative to the wings. However the kurtosis parameter, W50/W20, is not associated with the luminosity eigenvector.

There are other possibilities: Perhaps more luminous disks exist in deeper gravitational potentials and the NLR dynamics are gravitational, or more luminous disks have higher mass transfer rates, higher angular momentum, and radio jets with a stronger effect on accelerating NLR gas. Whittle (1992a, b, c) has shown that both gravitational and jet-related effects are important to the dynamics of the NLR in Seyfert galaxies. One possible conclusion is simply that more massive host galaxies (with broader [O III] $\lambda 5007$) have more massive central engines (that are more luminous). Further investigation of quasar host galaxies, jet-NLR interactions, and the differences between radio-loud and radio-quiet AGNs should help to resolve these issues.

## 5. Summary

The H$\beta$ and [O III] $\lambda 5007$ profiles have been investigated in a data set of 60 radio-loud quasars with similar numbers of core- and lobe-dominated objects, formed from the combination of new spectra from McDonald Observatory, the spectra of the radio-loud BQS sources from Boroson & Green (1992), and the high-resolution spectra of Jackson & Browne (1991).

Several statistical analyses have been performed, including principal component analyses (PCA). Three main effects are:

1. A number of H$\beta$ profile parameters are correlated with the core dominance, a measure of orientation in the relativistic beaming model. The most straightforward interpretation is that a very broad ($\sim$8000-9000 km s$^{-1}$), redshifted component is strong in lobe-dominant sources and weak in core-dominant sources. The strength of this component is inversely correlated with the Fe II/[O III] intensity ratio.

2. The second most significant source of variation in the H$\beta$ profile appears to involve an intermediate-width ($\sim$2000 km s$^{-1}$) component with a small redshift. The strength of this component is correlated with the strength of [O III] $\lambda 5007$.

3. The third effect depends on optical luminosity. As luminosity increases, $\alpha_{ox}$ increases, [O III] $\lambda 5007$ line width increases, and EW(H$\beta$) and EW([O III]) decrease (Baldwin effects).

These sets of correlations have been discussed in terms of other results in the literature and possible physical interpretations. The simple projection schemes (e.g., Wills & Browne 1986) used to explain the inverse correlation of core dominance and line width are not the most likely according to the results of PCA. Two significant non-correlations are noted for this sample contrary to previous studies: between FWHM$_{H\beta}$ and FWHM$_{5007}$, and between the asymmetry of [O III] $\lambda 5007$ and the core dominance. A quantitative comparison of this sample with radio-quiet objects is deferred to a later paper (Brotherton 1995).

I thank Bev and Derek Wills for assistance making observations as well as useful discussions and reading the manuscript of this paper. Many thanks also to David Doss, Doug Otoupal, Mike Ward, Marian Frueh, and the staff at McDonald Observatory for their invaluable assistance on many observing runs. I also gratefully acknowledge Bev Wills, Dongping Fang, Todd Boroson, Richard Green, Neal Jackson, and Ian Browne for making their archival data available in digital form. I thank Keith Thompson for providing his convolution programs. The referee made several good suggestions which improved this paper. I acknowledge support from a University of Texas Graduate Fellowship. Partial support for this work was provided by NASA through grant number GO-2578 from the Space Telescope Science Institute, which is operated by the Associate of Universities for Research in Astronomy, Inc., under NASA contract NAS5-26555.

**Figure Captions**

FIG. 1. (a-h) Final reduced spectra of the the 41 radio-loud quasars observed at McDonald Observatory. The vertical axis shows observed flux in units of $10^{-16}$ ergs cm$^{-2}$ s$^{-1}$ Å$^{-1}$. The bottom horizontal axis is the observed wavelength in Å, the top is the rest frame wavelength.

FIG. 2. Spectra of 0955+326 at two epochs. The top spectrum is from 1980 (courtesy B. Wills), with low resolution. The bottom spectrum is the new 1992 observation, multiplied by a factor to match the flux in [O III] $\lambda$5007. For both spectra, an estimate of the contribution of H$\beta_n$ is shown based on the profile of the [O III] $\lambda$5007.

FIG. 3. The distributions of the continuum and emission line properties given in Table 5. The core-dominated sample is denoted by a dotted line, while the lobe-dominated sample is denoted by a solid line.

FIG. 4. The correlations between line widths and the core dominance, using both R and R$_V$. All widths are in km s$^{-1}$. In all plots core-dominant objects are denoted by filled circles, lobe-dominant by open circles. The narrow-lined outlier, 0955+326, would fall on the trend using the 1980 epoch data. The broad-lined outlier is 0710+118. Only the upper right panel does not show a statistically significant correlation.

FIG. 5. Log of the rest frame EW of [O III] $\lambda$5007 in Å plotted against the log of the core dominance, R. Note the presence of a correlation among the core-dominant objects taken alone, but not for the lobe-dominant objects.

FIG. 6. Significant correlations involving EW$_{H\beta}$ (in rest frame Å). Line widths are in km s$^{-1}$. In the left panel, the outlier 0955+326, in the lower left hand corner, would move along the direction of the relation plotted using the 1980 epoch data.

FIG. 7. The correlations between line shape, asymmetry, and FW1/4M for H$\beta$. Larger values of Shape2 correspond to larger differences between the widths at the 1/4 and 3/4 levels. Asymm$_{H\beta}$ < 0 corresponds to a red wing stronger than the blue. The broadest-lined object, 0710+118, has more emission to the red of zero velocity as defined by [O III] $\lambda$5007, but excess blue emission if the line peak is taken to represent zero velocity. A different measurement of the asymmetry would put this object in line with other objects in the right panel.

FIG. 8. Significant correlations involving $\alpha_{ox}$, the optical to X-ray spectral index. IPV20 is in km s$^{-1}$, M$_{OIII}$ is in magnitudes, and positive values of AI20 indicate excess emission in the blue wing compared with the red.

FIG. 9. The correlation between the line width W50 in km s$^{-1}$ and M$_{abs}$.

FIG. 10. The intensity ratio of narrow [O III] $\lambda$5007 to narrow H$\beta$ (estimated as described in the text) plotted against $\alpha_{ox}$, the optical-to-X-ray spectral index.

FIG. 11. The top two panels show continuum-normalized mean spectra for the objects with FWHM$_{H\beta}$ less than and greater than <FWHM$_{H\beta}$>. The bottom panel is the difference.

TABLE 1
Data from the Literature

| Name | $M_{abs}$[a] (mag.) | S.I.[a] | Log R[b] | Log $R_V$ | $\alpha_{ox}$[c] | Log $L_{core}$ (W Hz$^{-1}$) | Log $L_{ext}$ (W Hz$^{-1}$) |
|---|---|---|---|---|---|---|---|
| 0003+158 | −26.0 | 0.59 | −0.38 | 1.95 | 1.34 | 26.04 | 26.42 |
| 0007+106 | −23.3 | −0.98 | 1.44 | 2.11 | 1.05 | 25.12 | 23.68 |
| 0044+030 | −27.2 | 1.36 | −0.42 | 0.83 | 1.59 | 25.40 | 25.82 |
| 0056−001 | −26.2 | 0.45 | >0.70 | 3.28 | 1.34 | 27.45 | <26.75 |
| 0115+027 | −26.5 | 0.85 | −0.50 | 2.25 | ⋯ | 26.54 | 27.04 |
| 0134+329 | −25.7 | 0.83 | −1.10 | 2.50 | 1.27 | 26.47 | 27.57 |
| 0159−117 | −27.1 | 0.60 | 0.16 | 2.75 | ⋯ | 27.28 | 27.12 |
| 0405−123 | −28.4 | 0.28 | 0.90 | 2.36 | 1.48 | 27.41 | 26.51 |
| 0414−060 | −27.8 | 0.82 | −0.50 | 1.68 | 1.44 | 26.49 | 26.99 |
| 0420−014 | −26.4 | 0.51 | 1.86 | 3.52 | 1.06 | 27.77 | 25.91 |
| 0710+118 | −27.1 | 1.01 | −1.85 | 1.09 | 1.58 | 25.62 | 27.47 |
| 0736+017 | −23.8 | 0.10 | 2.10 | 3.24 | 1.36 | 26.45 | 24.35 |
| 0738+313 | −27.2 | −0.23 | 1.27 | 3.04 | 1.58 | 27.61 | 26.34 |
| 0742+318 | −26.5 | 0.21 | 0.50 | 2.56 | ⋯ | 26.85 | 26.35 |
| 0837−120 | −24.7 | 0.56 | −0.70 | 1.80 | 1.24 | 25.37 | 26.07 |
| 0838+133 | −25.4 | 0.34 | −0.50 | 3.16 | 1.20 | 27.01 | 27.51 |
| 0903+169 | −24.5 | 0.87 | −1.40 | 1.70 | 1.13 | 25.19 | 26.59 |
| 0923+392 | −25.7 | −0.83 | 1.20 | 4.22 | 1.10 | 28.19 | 26.99 |
| 0952+097 | −24.1 | 0.72 | <−0.50 | <1.90 | ⋯ | <25.23 | 25.73 |
| 0955+326 | −27.1 | 0.28 | −0.07 | 2.23 | ≥1.65 | 26.76 | 26.83 |
| 1004+130 | −25.7 | 0.68 | −1.68 | 0.43 | ≥1.84 | 24.40 | 26.08 |
| 1007+417 | −27.0 | 0.62 | −0.39 | 2.17 | ⋯ | 26.66 | 27.05 |
| 1011+232 | −25.5 | −0.15 | 0.88 | 3.12 | ⋯ | 27.01 | 26.13 |
| 1020−103 | −24.2 | 0.62 | 0.66 | 2.34 | 1.32 | 25.71 | 25.05 |
| 1028+313 | −23.4 | 0.33 | 0.29 | 2.11 | 1.10 | 25.16 | 24.87 |
| 1048−090 | −24.9 | 0.92 | −1.31 | 1.66 | 1.34 | 25.31 | 26.62 |
| 1100+772 | −25.8 | 0.95 | −0.91 | 1.63 | 1.32 | 25.64 | 26.55 |
| 1103−006 | −25.8 | 0.78 | −0.14 | 2.20 | ⋯ | 26.21 | 26.35 |
| 1137+660 | −27.1 | 0.83 | −1.07 | 1.83 | 1.30 | 26.36 | 27.43 |
| 1146−037 | −24.8 | 0.48 | 0.16 | 2.36 | 1.16 | 25.97 | 25.81 |
| 1149−110 | −21.9 | ⋯ | −0.09 | ⋯ | ⋯ | ⋯ | ⋯ |
| 1150+497 | −24.6 | 0.57 | 0.13 | 2.99 | ⋯ | 26.52 | 26.39 |
| 1156+295 | ⋯ | 0.42 | 0.90 | ⋯ | ⋯ | 27.29 | 26.39 |
| 1217+023 | −24.3 | −0.03 | 0.11 | 2.43 | 1.12 | 25.84 | 25.73 |
| 1226+023 | −27.0 | −0.08 | 0.80 | 3.10 | 1.31 | 27.59 | 26.79 |
| 1250+568 | −23.6 | 0.61 | −1.62 | 2.01 | 1.34 | 25.14 | 26.76 |
| 1302−102 | −26.0 | −0.06 | 0.49 | 2.45 | 1.45 | 26.54 | 26.05 |
| 1305+069 | −26.1 | 1.12 | <−1.20 | <1.52 | ⋯ | <25.65 | 26.85 |
| 1309+355 | −24.6 | ⋯ | ⋯ | ⋯ | ⋯ | ⋯ | ⋯ |
| 1340+290 | −27.0 | 0.00 | ⋯ | ⋯ | ⋯ | ⋯ | ⋯ |
| 1351+267 | −24.3 | 0.63 | −0.30 | 1.72 | ⋯ | 25.13 | 25.43 |
| 1354+195 | −27.5 | 0.24 | 0.37 | 2.73 | ⋯ | 27.42 | 27.05 |
| 1425+267 | −26.2 | 0.42 | −0.40 | 1.03 | 1.69 | 25.20 | 25.60 |
| 1458+718 | −27.3 | 0.56 | −1.00 | 2.74 | 1.33 | 27.35 | 28.35 |
| 1510−089 | −25.3 | −0.78 | 1.31 | 3.54 | 1.29 | 27.35 | 26.04 |
| 1512+370 | −25.6 | 0.14 | −0.71 | 1.88 | 1.46 | 25.81 | 26.52 |
| 1522+155 | −25.8 | 0.38 | 0.90 | 2.72 | 1.50 | 26.73 | 25.83 |
| 1545+210 | −24.4 | 0.55 | −1.32 | 1.74 | 1.25 | 25.19 | 26.51 |
| 1618+177 | −26.5 | 1.36 | −0.67 | 1.98 | 1.49 | 26.27 | 26.94 |
| 1622+238 | −26.7 | 1.05 | −1.72 | 1.63 | ⋯ | 26.00 | 27.72 |
| 1641+399 | −27.1 | 0.10 | 1.23 | 3.38 | ⋯ | 27.91 | 26.68 |
| 1656+053 | −27.4 | −0.44 | 0.61 | 3.16 | 1.26 | 27.81 | 27.20 |
| 1704+608 | −26.6 | 0.84 | −1.91 | 0.73 | 1.56 | 25.06 | 26.97 |
| 1721+343 | −24.0 | 0.18 | 0.00 | 2.72 | 0.98 | 26.01 | 26.01 |
| 1750+175 | −27.2 | ⋯ | 1.30 | 2.05 | ⋯ | 26.62 | 25.32 |
| 2135−147 | −24.9 | 0.75 | −0.70 | 1.97 | 1.30 | 25.62 | 26.32 |
| 2209+184 | −22.3 | ⋯ | 1.56 | ⋯ | 1.26 | ⋯ | ⋯ |
| 2251+113 | −25.8 | 0.35 | −1.52 | 1.00 | 1.70 | 25.01 | 26.53 |
| 2308+098 | −26.3 | 0.59 | −0.78 | 1.41 | ⋯ | 25.62 | 26.40 |
| 2344+092 | −27.5 | 0.08 | 2.00 | 2.75 | 1.51 | 27.44 | 25.44 |

[a] From Véron-Cetty & Véron (1989), who use $H_o = 50$ km s$^{-1}$ Mpc$^{-1}$ and $q_o = 0$. S.I. is the radio spectral index based on 6 and 11 cm flux densities.

[b] These values are from Wills et al. (1992), Wills & Browne (1986) and references therein. R is the ratio of radio flux density in the core to the lobes at 5 GHz rest frame, a measure of core dominance.

[c] From Wilkes et al. (1994), based on *Einstein* data. They use $H_o = 50$ km s$^{-1}$ Mpc$^{-1}$ and $q_o = 0$, and have assumed an X-ray slope of $\alpha_x = 0.5$ to calculate X-ray luminosities.

TABLE 2
LOG OF NEW MCDONALD SPECTROSCOPY

| Target | $z^a$ | $V^a$ (mag.) | Date[b] (UT) | Grat.[c] | Slit ($''$) | Exp-T (sec) |
|---|---|---|---|---|---|---|
| 0056−001 | 0.717 | 17.3 | 1992 Nov 29 | g44 | 2 | 6900 |
| 0115+027 | 0.670 | 17.0 | 1992 Nov 30 | g44 | 2 | 8100 |
| 0159−117 | 0.670 | 16.4 | 1992 Nov 29 | g44 | 2 | 7200 |
| 0414−060 | 0.773 | 15.9 | 1993 Feb 20 | g44 | 2 | 2700 |
|  |  |  | 1993 Feb 21 | g44 | 2 | 5400 |
| 0420−014 | 0.915 | 17.0 | 1993 Jan 16 | g45 | 2 | 8100 |
|  |  |  | 1993 Feb 23 | g45 | 2 | 2700 |
| 0710+118 | 0.768 | 16.6 | 1993 Jan 20 | g44 | 2 | 2700 |
|  |  |  | 1993 Feb 22 | g44 | 2 | 5400 |
|  |  |  | 1993 Feb 23 | g44 | 2 | 2700 |
| 0736+017 | 0.191 | 16.5 | 1992 Nov 27 | g44 | 2 | 5400 |
| 0738+313 | 0.630 | 16.1 | 1992 Nov 29 | g44 | 2 | 5400 |
|  |  |  | 1992 Nov 30 | g44 | 2 | 1800 |
| 0742+318 | 0.462 | 16.0 | 1992 Feb 29 | g44 | 2 | 2400 |
|  |  |  | 1992 Mar 02 | g44 | 2 | 2400 |
| 0837−120 | 0.200 | 15.8 | 1990 Jan 29 | g44 | 2.5 | 5400 |
| 0838+133 | 0.684 | 18.1 | 1992 Nov 30 | g44 | 2 | 13500 |
| 0903+169 | 0.411 | 18.3 | 1992 Feb 28 | g44 | 2 | 8100 |
|  |  |  | 1992 Feb 29 | g44 | 2 | 5400 |
| 0923+392 | 0.698 | 17.9 | 1992 Nov 29 | g44 | 2 | 7200 |
| 0952+097 | 0.298 | 17.2 | 1992 Nov 27 | g44 | 2 | 4500 |
| 0955+326 | 0.533 | 15.8 | 1992 Mar 02 | g44 | 2 | 2700 |
| 1004+130 | 0.240 | 15.2 | 1990 Jan 30 | g44 | 2.5 | 3600 |
| 1007+417 | 0.613 | 16.2 | 1992 Mar 02 | g44 | 2 | 5400 |
| 1011+232 | 0.565 | 17.5 | 1993 Apr 21 | g44 | 2 | 7200 |
| 1020−103 | 0.197 | 16.1 | 1992 Mar 01 | g44 | 2 | 2700 |
| 1028+313 | 0.177 | 16.7 | 1992 Mar 01 | g44 | 2 | 4500 |
| 1100+772 | 0.311 | 15.7 | 1990 Apr 24 | g44 | 3 | 4800 |
| 1103−006 | 0.426 | 16.5 | 1992 Feb 29 | g44 | 2 | 5100 |
| 1137+660 | 0.652 | 16.3 | 1992 Jun 01 | g44 | 2 | 3600 |
|  |  |  | 1992 Jun 07 | g44 | 2 | 3600 |
| 1146−037 | 0.341 | 16.9 | 1992 Feb 29 | g44 | 2 | 4500 |
| 1150+497 | 0.334 | 17.1 | 1992 Feb 28 | g44 | 2 | 5400 |
| 1156+295 | 0.729 | 14.4 | 1993 Apr 20 | g44 | 2 | 3600 |
|  |  |  | 1993 Apr 22 | g42 | 2 | 2700 |
| 1217+023 | 0.240 | 16.5 | 1993 Apr 21 | g44 | 2 | 3600 |
| 1250+568 | 0.320 | 17.9 | 1992 Feb 28 | g44 | 2 | 2700 |
|  |  |  | 1992 Feb 29 | g44 | 2 | 5400 |
| 1305+069 | 0.599 | 17.0 | 1992 Mar 02 | g44 | 2 | 6000 |
| 1340+290 | 0.905 | 17.1 | 1994 Apr 09 | g42 | 2 | 7200 |
|  |  |  | 1994 Apr 10 | g42 | 2 | 2700 |
| 1354+195 | 0.720 | 16.0 | 1993 Apr 20 | g44 | 2 | 4500 |
| 1458+718 | 0.905 | 16.8 | 1993 Apr 22 | g42 | 2 | 2400 |
|  |  |  | 1994 Apr 08 | g42 | 2 | 2400 |
|  |  |  | 1994 Apr 10 | g42 | 2 | 2400 |
| 1510−089 | 0.360 | 16.5 | 1992 Feb 28 | g44 | 2 | 2700 |
| 1522+155 | 0.628 | 17.5 | 1993 Apr 20 | g44 | 2 | 5400 |
| 1545+210 | 0.264 | 16.7 | 1990 Apr 24 | g44 | 3 | 2400 |
| 1618+177 | 0.555 | 16.4 | 1992 Jun 01 | g44 | 2 | 5400 |
| 1622+238 | 0.927 | 17.5 | 1994 Apr 09 | g42 | 2 | 4500 |
|  |  |  | 1994 Apr 10 | g42 | 2 | 8100 |
| 1641+399 | 0.594 | 16.0 | 1992 Jun 01 | g44 | 2 | 3600 |
|  |  |  | 1992 Jun 07 | g44 | 2 | 3600 |
| 1656+053 | 0.879 | 16.5 | 1993 Apr 22 | g42 | 2 | 4800 |
| 1750+175 | 0.507 | 15.5 | 1993 Apr 21 | g44 | 2 | 2400 |
| 2344+092 | 0.673 | 16.0 | 1992 Nov 30 | g44 | 2 | 5400 |

[a] From Véron-Cetty & Véron (1989).
[b] Observations made before 1992 April used a TI 800 × 800 thinned CCD, those made after 1992 April used a thick Craf-Cassini 1024 × 1024 CCD.
[c] Grating g42 has 300 l/mm and is blazed at 1$\mu$m. Gratings g44 and g45 have 600 l/mm and are blazed at 7500Å and 1$\mu$m, respectively. With a $2''$ slit, the 300 l/mm grating gives $\sim 10$Å resolution and $\sim 2800$Å coverage, while the 600 l/mm gratings give $\sim 5$ Å resolution and 1400Å coverage.

TABLE 3
Emission Line Measurements[a]

| Name | EW Hβ (Å) | FW1/4M Hβ (km s$^{-1}$) | FWHM Hβ (km s$^{-1}$) | FW3/4M Hβ (km s$^{-1}$) | Shift Hβ | Shape Hβ | Shape2 Hβ | Asymm Hβ | EW λ5007 (Å) | EW[b] Fe II (Å) | Flag[c] | He II[d] λ4686 |
|---|---|---|---|---|---|---|---|---|---|---|---|---|
| 0003+158[e] | 91 | 8720 | 4760 | 2160 | −0.05 | 1.14 | 1.38 | −0.16 | 26 | 0 | ... | b |
| 0007+106[e] | 101 | 7570 | 5100 | 3140 | 0.02 | 1.05 | 0.87 | −0.05 | 42 | 35 | ... | n |
| 0044+030[f] | 55 | 7100 | 5100 | 2900 | 0.00 | 0.98 | 0.82 | −0.02 | 21 | 21[g] | B | ... |
| 0056−001 | 28 | 4300 | 3000 | 1900 | 0.03 | 1.03 | 0.80 | 0.02 | 50 | 0 | C | b: |
| 0115+027 | 83 | 7400 | 5000 | 3100 | 0.07 | 1.05 | 0.86 | −0.15 | 47 | 0 | C | ... |
| 0134+329[f] | 45 | 5200 | 3800 | 2700 | 0.05 | 1.04 | 0.66 | 0.14 | 45 | 67 | B | ... |
| 0159−117 | 64 | 6300 | 4500 | 2600 | −0.02 | 0.99 | 0.82 | 0.01 | 19 | 80 | B | ... |
| 0405−123[f] | 57 | 8800 | 4800 | 2700 | 0.04 | 1.21 | 1.27 | −0.27 | 12 | 20 | B | ... |
| 0414−060 | 207 | 15600 | 8200 | 4500 | 0.08 | 1.23 | 1.35 | −0.32 | 25 | 0 | B | ... |
| 0420−014 | 43 | ... | 3000 | ... | ... | ... | ... | ... | ... | 0 | D | ... |
| 0710+118 | 90 | 25000 | 20000 | 12000 | 0.20 | 0.93 | 0.65 | −0.17 | 22 | 0 | B | ... |
| 0736+017 | 51 | 4400 | 3400 | 2100 | 0.09 | 0.96 | 0.68 | 0.03 | 5 | 24 | A | ... |
| 0738+313 | 108 | 7800 | 4800 | 3000 | −0.01 | 1.13 | 1.00 | −0.08 | 17 | 47 | B | ... |
| 0742+318 | 95 | 14500 | 9940 | 5310 | 0.02 | 1.00 | 0.92 | −0.19 | 41 | 0 | A | n |
| 0837−120 | 180 | 15000 | 6060 | 2780 | 0.02 | 1.48 | 2.02 | −0.22 | 54 | 36 | A | n:, b |
| 0838+133 | 46 | 3700 | 3000 | 1800 | 0.17 | 0.92 | 0.63 | −0.07 | 21 | 0 | B | b |
| 0903+169 | 34 | 6300 | 4440 | 1600 | −0.05 | 0.89 | 1.06 | −0.03 | 53 | 0 | A | n, b |
| 0923+392 | 64 | 12500 | 7200 | 3800 | 0.19 | 1.32 | 1.21 | −0.26 | 13 | 0 | B | ... |
| 0952+097 | 80 | 6500 | 3800 | 2100 | −0.07 | 1.13 | 1.16 | 0.07 | 120 | 0 | C | n |
| 0955+326 | 12 | 3090 | 1380 | 620 | −0.06 | 1.34 | 1.79 | −0.31 | 48 | 17 | A | ... |
| 1004+130[e] | 43 | 13010 | 6300 | 4061 | 0.17 | 1.35 | 1.42 | 0.07 | 6 | 10 | A | ... |
| 1007+417 | 43 | 5370 | 3560 | 1910 | 0.01 | 1.02 | 0.97 | 0.09 | 21 | 0 | B | b |
| 1011+232 | 52 | 4100 | 2700 | 1400 | −0.04 | 1.02 | 1.00 | −0.13 | 37 | 22 | B | ... |
| 1020−103 | 128 | 12100 | 9750 | 6170 | −0.07 | 0.94 | 0.61 | −0.19 | 39 | 47 | A | b: |
| 1028+313 | 67 | 10060 | 5000 | 1850 | 0.08 | 1.19 | 1.64 | −0.21 | 38 | 0 | A | n |
| 1048−090[e] | 81 | 11030 | 5620 | 2660 | 0.07 | 1.22 | 1.49 | −0.22 | 28 | 0 | ... | ... |
| 1100+772[e] | 90 | 10840 | 6160 | 2800 | 0.06 | 1.11 | 1.31 | −0.10 | 41 | 19 | A | n |
| 1103−006 | 62 | 10000 | 6560 | 3330 | −0.04 | 1.02 | 1.02 | −0.18 | 7 | 40 | A | b: |
| 1137+660 | 72 | 9690 | 6060 | 3520 | 0.06 | 1.09 | 1.02 | −0.15 | 17 | 0 | B | ... |
| 1146−037 | 112 | 11110 | 5500 | 2470 | −0.01 | 1.23 | 1.57 | −0.20 | 316 | 0 | C | ... |
| 1149−110[e] | 118 | 4920 | 3060 | 1900 | −0.02 | 1.11 | 0.99 | −0.13 | 33 | 43 | ... | n, b |
| 1150+497 | 52 | 7160 | 4810 | 2900 | 0.02 | 1.05 | 0.89 | −0.12 | 21 | 19 | A | n |
| 1156+295 | 10 | ... | 3700 | ... | ... | ... | ... | ... | 6 | ... | D | ... |
| 1217+023 | 100 | 7200 | 4300 | 2000 | −0.02 | 1.07 | 1.21 | −0.15 | 39 | 0 | A | b |
| 1226+023[e] | 113 | 6580 | 3520 | 1960 | 0.04 | 1.21 | 1.31 | 0.04 | 5 | 64 | ... | ... |
| 1250+568 | 100 | 6790 | 4560 | 2720 | 0.05 | 1.04 | 0.89 | −0.07 | 150 | 0 | A | n, b |
| 1302−102[e] | 28 | 5080 | 3400 | 1860 | 0.03 | 1.02 | 0.95 | −0.02 | 9 | 17 | ... | ... |
| 1305+069 | 60 | 15100 | 6440 | 3090 | −0.10 | 1.41 | 1.86 | −0.26 | 33 | 0 | B | b: |
| 1309+355[e] | 51 | 9640 | 2940 | 1660 | 0.04 | 1.92 | 2.71 | −0.08 | 19 | 14 | ... | ... |
| 1340+290 | 170 | 18600 | 13000 | 5000 | 0.00 | 0.91 | 1.05 | −0.12 | 36 | 23 | B | b |
| 1351+267[f] | 128 | 11700 | 8600 | 2700 | 0.01 | 0.98 | 0.77 | −0.13 | 48 | 0 | B | n |
| 1354+195 | 75 | 7200 | 4400 | 2300 | 0.06 | 1.08 | 1.11 | −0.03 | 73 | 0 | B | b |
| 1425+267[e] | 93 | 17960 | 9410 | 4700 | 0.05 | 1.20 | 1.41 | −0.05 | 36 | 10 | ... | n |
| 1458+718 | 61 | 4400 | 3000 | 1600 | 0.37 | 1.00 | 0.93 | −0.43 | 86 | 0 | B | b |
| 1510−089 | 56 | 4130 | 2880 | 1790 | 0.21 | 1.03 | 0.81 | 0.02 | 8 | 84 | A | ... |
| 1512+370[e] | 122 | 13330 | 6810 | 3710 | 0.02 | 1.25 | 1.41 | −0.04 | 57 | 0 | ... | b |
| 1522+155 | 30 | 6800 | 4600 | 3300 | 0.16 | 1.10 | 0.76 | 0.13 | 27 | 0 | C | ... |
| 1545+210[e] | 96 | 9380 | 7030 | 4590 | 0.09 | 0.99 | 0.68 | −0.15 | 33 | 0 | A | n, b |
| 1618+177 | 45 | 10370 | 7000 | 2840 | 0.02 | 0.94 | 1.08 | −0.01 | 37 | 0 | B | ... |
| 1622+238 | 104 | 12000 | 7100 | 2800 | −0.04 | 1.04 | 1.30 | −0.27 | 42 | 0 | D | ... |
| 1641+399 | 10 | 6050 | 3560 | 2350 | 0.06 | 1.18 | 1.04 | −0.19 | 2 | 4 | B | ... |
| 1656+053 | 51 | ... | 5000 | ... | ... | ... | ... | ... | ... | 87 | D | ... |
| 1704+608[e] | 28 | 13080 | 6560 | 3710 | 0.04 | 1.28 | 1.43 | −0.29 | 27 | 0 | ... | n |
| 1721+343[f] | 95 | 8400 | 2600 | 1400 | −0.08 | 1.88 | 2.69 | −0.23 | 43 | 0 | A | n, b |
| 1750+175 | 65 | 5200 | 3700 | 2200 | 0.04 | 1.00 | 0.81 | 0.04 | 19 | 0 | A | b: |
| 2135−147[f] | 67 | 10000 | 7300 | 3700 | 0.04 | 0.95 | 0.87 | −0.08 | 61 | 0 | C | ... |
| 2209+184[e] | 115 | 10500 | 6500 | 4990 | −0.07 | 1.19 | 0.85 | 0.05 | 15 | 50 | ... | ... |
| 2251+113[e] | 82 | 7920 | 4160 | 2200 | −0.13 | 1.22 | 1.38 | −0.08 | 19 | 26 | ... | n |
| 2308+098[e] | 90 | 14690 | 7970 | 3570 | −0.01 | 1.15 | 1.40 | −0.19 | 17 | 0 | ... | b |
| 2344+092 | 85 | 7600 | 3900 | 2600 | 0.01 | 1.31 | 1.28 | −0.19 | 61 | 0 | A | b: |

[a] See § 2.3.1 for parameter definitions.
[b] The EWs were derived from the Fe II template between λ4434 and λ4684 and are not direct measurements.
[c] Quality flags. Class "A" spectra, have SNR > 20 per pixel and errors of ∼5% for basic line measurements. Class "D" spectra have SNR ∼ 5 per pixel, errors of ∼20%, and are unsuitable for most measurements. Classes "B" and "C" are intermediate cases.
[d] Reports the presence of narrow ("n") and broad ("b") He II λ4686. A colon indicates uncertainty.
[e] The measurements are from Boroson & Green (1992). Their spectra are class "A" and "B."
[f] The profiles were measured from the high-resolution spectra of Jackson & Browne (1991).
[g] Estimated from Grandi (1981).

TABLE 4
[O III] $\lambda$5007 Profile Measurements[a,b]

| Name | EW (Å) | $M_{OIII}$ (mag.) | [OIII]/H$\beta$ | W50 ($km\ s^{-1}$) | W20 ($km\ s^{-1}$) | W50/W20 | AI20 | IPV20 ($km\ s^{-1}$) | Asy20 | FWHM$_i$ ($km\ s^{-1}$) |
|---|---|---|---|---|---|---|---|---|---|---|
| 0044+030[c] | 21 | −30.5 | 0.38 | 820 | 1300 | 0.63 | 0.39 | 1170 | 0.27 | 90 |
| 0056−001 | 50 | −30.4 | 1.79 | 980 | 1480 | 0.67 | 0.00 | 1160 | −0.02 | 170 |
| 0115+027 | 47 | −30.7 | 0.57 | 680 | 1170 | 0.58 | 0.25 | 1430 | 0.00 | 180 |
| 0134+329[c] | 45 | −29.8 | 1.00 | 1200 | 1640 | 0.73 | 0.35 | 1430 | 0.58 | 110 |
| 0159−117 | 19 | −30.3 | 0.30 | 620 | 1480 | 0.42 | 0.40 | 1240 | 0.65 | 180 |
| 0405−123[c] | 12 | −31.1 | 0.21 | 710 | 1800 | 0.39 | 0.09 | 1220 | 0.20 | 95 |
| 0414−060 | 25 | −31.3 | 0.12 | 580 | 1020 | 0.56 | −0.14 | 820 | −0.16 | 170 |
| 0710+118 | 22 | −30.5 | 0.24 | 420 | 760 | 0.55 | 0.12 | 760 | 0.43 | 170 |
| 0736+017 | 5 | −25.5 | 0.10 | 540 | 1190 | 0.46 | 0.36 | ... | ... | 250 |
| 0738+313 | 17 | −30.3 | 0.16 | 360 | 1220 | 0.29 | 0.52 | 1000 | 0.83 | 180 |
| 0742+318 | 41 | −30.5 | 0.43 | 460 | 740 | 0.62 | 0.12 | 560 | 0.82 | 200 |
| 0837−120 | 54 | −29.0 | 0.30 | 520 | 1170 | 0.45 | −0.29 | 1090 | −0.19 | 300 |
| 0838+133 | 21 | −28.7 | 0.46 | 360 | 590 | 0.60 | 0.08 | 880 | −0.07 | 180 |
| 0903+169 | 53 | −28.8 | 1.54 | 510 | 1000 | 0.51 | 0.14 | 810 | 0.50 | 210 |
| 0923+392 | 13 | −28.5 | 0.20 | 520 | 750 | 0.69 | −0.12 | 690 | 0.00 | 180 |
| 0952+097 | 120 | −29.3 | 1.50 | 330 | 490 | 0.67 | 0.00 | 410 | 0.47 | 230 |
| 0955+326 | 48 | −31.3 | 3.93 | 880 | 1420 | 0.62 | 0.17 | 1300 | 0.29 | 200 |
| 1004+130 | 6 | −27.6 | 0.14 | 470 | 1060 | 0.44 | 0.30 | 970 | 0.51 | 290 |
| 1007+417 | 21 | −30.3 | 0.48 | 520 | 1120 | 0.46 | −0.04 | 820 | 0.16 | 190 |
| 1011+232 | 37 | −29.4 | 0.71 | 350 | 530 | 0.67 | 0.07 | 500 | 0.47 | 190 |
| 1020−103 | 39 | −28.2 | 0.30 | 430 | 870 | 0.50 | 0.05 | 730 | 0.30 | 250 |
| 1028+313 | 38 | −27.3 | 0.56 | 190 | 400 | 0.49 | −0.03 | 390 | 0.00 | 250 |
| 1100+772 | 41 | −29.8 | 0.46 | 610 | 940 | 0.65 | 0.05 | 740 | 0.29 | 230 |
| 1103−006 | 7 | −27.8 | 0.11 | 770 | 1210 | 0.64 | 0.28 | 930 | 0.56 | 210 |
| 1137+660 | 17 | −30.2 | 0.24 | 570 | 750 | 0.76 | 0.12 | 630 | 0.33 | 180 |
| 1146−037 | 316 | −31.0 | 2.82 | 330 | 500 | 0.67 | 0.00 | 450 | 0.29 | 220 |
| 1150+497 | 21 | −27.9 | 0.40 | 390 | 610 | 0.64 | 0.14 | 410 | 0.90 | 220 |
| 1156+295 | 6 | ... | 0.60 | 360 | 540 | 0.67 | −0.07 | 540 | −0.20 | 350 |
| 1217+023 | 39 | −28.3 | 0.39 | 380 | 660 | 0.58 | −0.07 | 540 | 0.00 | 240 |
| 1250+568 | 150 | −29.0 | 1.50 | 500 | 830 | 0.60 | 0.00 | 580 | 0.34 | 230 |
| 1305+069 | 33 | −29.9 | 0.55 | 670 | 1010 | 0.67 | 0.14 | 790 | 0.37 | 190 |
| 1340+290 | 36 | −30.9 | 0.21 | 450 | 770 | 0.59 | 0.33 | 770 | 0.83 | 310 |
| 1351+267[c] | 48 | −28.5 | 0.38 | 280 | 650 | 0.43 | −0.04 | 600 | −0.21 | 115 |
| 1354+195 | 73 | −32.2 | 0.97 | 470 | 760 | 0.62 | 0.00 | 510 | 0.15 | 170 |
| 1425+267[c] | 36 | −30.1 | 0.39 | 310 | 730 | 0.42 | 0.17 | 780 | 0.23 | 110 |
| 1458+718 | 86 | −32.1 | 1.41 | 680 | 1100 | 0.62 | −0.08 | 770 | −0.21 | 310 |
| 1510−089 | 8 | −27.6 | 0.14 | 450 | 1360 | 0.33 | 0.57 | 930 | 1.00 | 220 |
| 1512+370[c] | 57 | −30.0 | 0.47 | 260 | 1070 | 0.24 | 0.63 | 560 | 0.73 | 110 |
| 1522+155 | 27 | −29.4 | 0.90 | 680 | 1060 | 0.63 | 0.00 | 810 | 0.16 | 180 |
| 1545+210 | 33 | −28.2 | 0.34 | 500 | 830 | 0.60 | 0.00 | 640 | 0.06 | 240 |
| 1618+177 | 37 | −30.4 | 0.82 | 720 | 1010 | 0.72 | 0.05 | 880 | 0.22 | 190 |
| 1622+238 | 42 | −30.8 | 0.40 | 390 | 650 | 0.60 | 0.00 | 830 | −0.28 | 310 |
| 1641+399 | 2 | −27.9 | 0.19 | 570 | 800 | 0.71 | 0.06 | 880 | −0.22 | 190 |
| 1704+608[c] | 27 | −30.2 | 0.96 | 410 | 640 | 0.64 | 0.16 | 1050 | 0.21 | 110 |
| 1721+343[c] | 43 | −28.1 | 0.45 | 380 | 730 | 0.52 | 0.07 | 740 | 0.08 | 120 |
| 1750+175 | 19 | −30.4 | 0.29 | 510 | 850 | 0.60 | 0.00 | 810 | 0.00 | 200 |
| 2135−147[c] | 61 | −29.4 | 0.91 | 310 | 570 | 0.54 | −0.10 | 530 | 0.10 | 120 |
| 2344+092 | 61 | −32.0 | 0.72 | 570 | 1220 | 0.47 | 0.28 | 1000 | 0.44 | 180 |

[a] See § 2.3.2 for parameter definitions.
[b] Line width measurements of narrow [O III] $\lambda$5007 have been corrected for instrumental resolution (FWHM$_i$); FWHM$_i$ and W50 are not significantly correlated indicating that line widths are not significantly influenced by resolution effects.
[c] The H$\beta$ line profile measurements are measured from the high-resolution digital spectra of Jackson & Browne (1991).

TABLE 5
MEAN PROPERTIES OF QUASAR SUBSAMPLES[a]

|  | All Quasars | Core-Dominant | Lobe-Dominant | Prob(W)[b] |
|---|---|---|---|---|
| $<z>$ | 0.47±0.03(60) | 0.48±0.05(26) | 0.46±0.04(32) | 0.80 |
| $<M_{abs}>$[c] (mag.) | −25.7±0.2(59) | −25.8±0.3(25) | −25.7±0.2(32) | 0.72 |
| $<\alpha_{ox}>$ | 1.36±0.03(41) | 1.29±0.04(18) | 1.41±0.04(24) | 0.09 |
| $<S.I.>$ | 0.43±0.07(56) | 0.07±0.09(24) | 0.71±0.05(31) | <0.01 |
| $<\text{Log } R>$ | −0.06±0.14(58) | 0.92±0.11(26) | −0.86±0.10(32) | <0.01 |
| $<\text{Log } R_V>$ | 2.24±0.11(55) | 2.84±0.11(25) | 1.75±0.11(30) | <0.01 |
| $<\text{Log } L_{core}>$[d] | 26.30±0.13(56) | 28.89±0.17(26) | 25.79±0.13(30) | <0.01 |
| $<\text{Log } L_{extended}>$[d] | 26.39±0.11(56) | 26.03±0.17(26) | 26.70±0.12(30) | <0.01 |
| $<\text{EW H}\beta>$ (Å) | 76±5(60) | 68±7(26) | 81±7(32) | 0.32 |
| $<\text{FW1/4M H}\beta>$ ($km\ s^{-1}$) | 9300±600(57) | 7700±600(23) | 10100±800(32) | 0.05 |
| $<\text{FWHM H}\beta>$ ($km\ s^{-1}$) | 5500±400(60) | 4700±400(26) | 6000±600(32) | 0.04 |
| $<\text{FW3/4M H}\beta>$ ($km\ s^{-1}$) | 3000±200(57) | 2800±300(23) | 3200±300(32) | 0.30 |
| $<\text{Shift H}\beta>$ | 0.03±0.01(57) | 0.04±0.02(23) | 0.03±0.02(32) | 0.78 |
| $<\text{Shape H}\beta>$ | 1.13±0.03(57) | 1.09±0.03(23) | 1.14±0.04(32) | 0.69 |
| $<\text{Shape2 H}\beta>$ | 1.16±0.06(57) | 1.02±0.06(23) | 1.21±0.08(32) | 0.08 |
| $<\text{Asymm H}\beta>$ | −0.11±0.02(57) | −0.08±0.02(23) | −0.12±0.02(32) | 0.29 |
| $<\text{Fe II/H}\beta>$[e] | 0.26±0.06(59) | 0.37±0.10(25) | 0.18±0.07(32) | 0.02 |
| $<\lambda 5007/\text{H}\beta>$ | 0.60±0.09(58) | 0.55±0.13(24) | 0.67±0.13(32) | 0.32 |
| $<\text{EW } \lambda 5007>$ (Å) | 40±6(58) | 38±13(24) | 41±5(32) | 0.05 |
| $<M_{[OIII]}\ \lambda 5007>$ (mag.) | −29.3±0.2(57) | −29.0±0.4(23) | −29.5±0.2(32) | 0.31 |
| $<W50\ \lambda 5007>$ ($km\ s^{-1}$) | 520±30(48) | 490±40(20) | 540±40(27) | 0.54 |
| $<W20\ \lambda 5007>$ ($km\ s^{-1}$) | 940±50(48) | 940±90(20) | 940±50(27) | 0.92 |
| $<W50/W20\ \lambda 5007>$ | 0.57±0.02(48) | 0.56±0.03(20) | 0.57±0.02(27) | 0.94 |
| $<AI20\ \lambda 5007>$ | 0.11±0.03(48) | 0.12±0.05(20) | 0.10±0.04(27) | 0.90 |
| $<IPV20\ \lambda 5007>$ ($km\ s^{-1}$) | 810±40(47) | 760±60(19) | 850±50(27) | 0.24 |
| $<Asy20\ \lambda 5007>$ | 0.27±0.05(46) | 0.32±0.09(18) | 0.18±0.05(27) | 0.52 |
| $<\lambda 5007/\text{H}\beta_n>$ | 11.0±0.7(52) | 10.1±0.9(20) | 11.2±0.8(30) | 0.37 |

[a] Entries in columns 2–4 are mean values, where the uncertainty is the error in the mean, with the number of objects following in parentheses. Refer to Tables 1–4 and § 2.3.1 and § 2.3.2 for additional information on each parameter.

[b] These are two-tailed probabilities that the medians (*not* the means) of the core- and lobe-dominant subsamples do not differ, computed using the non-parametric Wilcoxon rank sum test.

[c] Due to large variability, 1156+295 has not been included.

[d] The core and extended radio luminosities (5 GHz rest frame, in units of W Hz$^{-1}$) have been calculated from the R values in Table 1 and the 6 cm flux densities from Véron-Cetty & Véron (1989), assuming $H_o = 50$ km s$^{-1}$ Mpc$^{-1}$ and $q_o = 0$. A spectral index of 0 has been assumed for the core component, −1 for the extended component.

[e] Where Fe II was weak, upper limits were not estimated. Often this was true for the broader-lined objects, so this quantitative comparison is not exact.

TABLE 6
Rank–Correlation Matrix and Probabilities[a,b]

| | z | Mabs | $\alpha_{ox}$ | S.I. | R | $R_V$ | EW H$\beta$ | FW1/4M H$\beta$ | FWHM H$\beta$ | FW3/4M H$\beta$ | Shift H$\beta$ | Shape H$\beta$ | Shape2 H$\beta$ |
|---|---|---|---|---|---|---|---|---|---|---|---|---|---|
| Mabs | −0.77 | | | | | | | | | | | | |
|  | $< 10^{-6}$ | | | | | | | | | | | | |
| $\alpha_{ox}$ | 0.21 | −0.54 | | | | | | | | | | | |
|  | ... | $< 10^{-3}$ | | | | | | | | | | | |
| S.I. | 0.08 | −0.04 | 0.24 | | | | | | | | | | |
|  | ... | ... | ... | | | | | | | | | | |
| R | 0.01 | −0.03 | −0.24 | −0.73 | | | | | | | | | |
|  | ... | ... | ... | $< 10^{-6}$ | | | | | | | | | |
| $R_V$ | 0.20 | −0.05 | −0.42 | −0.67 | 0.77 | | | | | | | | |
|  | ... | ... | 0.001 | $< 10^{-6}$ | $< 10^{-6}$ | | | | | | | | |
| EW H$\beta$ | −0.27 | 0.21 | −0.19 | −0.04 | −0.10 | −0.27 | | | | | | | |
|  | 0.04 | ... | ... | ... | ... | 0.05 | | | | | | | |
| FW1/4M H$\beta$ | 0.01 | −0.03 | 0.10 | 0.30 | −0.36 | −0.61 | 0.51 | | | | | | |
|  | ... | ... | ... | 0.03 | 0.006 | $< 10^{-6}$ | $< 10^{-4}$ | | | | | | |
| FWHM H$\beta$ | 0.03 | −0.04 | 0.16 | 0.39 | −0.37 | −0.59 | 0.49 | 0.88 | | | | | |
|  | ... | ... | ... | 0.003 | 0.004 | $< 10^{-5}$ | $< 10^{-4}$ | $< 10^{-6}$ | | | | | |
| FW3/4M H$\beta$ | 0.09 | −0.08 | 0.25 | 0.29 | −0.19 | 0.45 | 0.41 | 0.76 | 0.91 | | | | |
|  | ... | ... | ... | 0.04 | ... | $< 10^{-3}$ | 0.002 | $< 10^{-6}$ | $< 10^{-6}$ | | | | |
| Shift H$\beta$ | 0.17 | −0.16 | 0.05 | −0.05 | −0.04 | 0.15 | −0.24 | −0.03 | 0.03 | 0.12 | | | |
|  | ... | ... | ... | ... | ... | ... | 0.08 | ... | ... | ... | | | |
| Shape H$\beta$ | −0.22 | 0.00 | 0.13 | −0.14 | −0.02 | −0.17 | 0.11 | 0.27 | −0.11 | −0.10 | −0.10 | | |
|  | 0.10 | ... | ... | ... | ... | ... | ... | 0.04 | ... | ... | ... | | |
| Shape2 H$\beta$ | −0.12 | −0.06 | 0.06 | −0.01 | −0.25 | −0.35 | 0.11 | 0.41 | 0.03 | −0.17 | −0.23 | 0.80 | |
|  | ... | ... | ... | ... | 0.06 | 0.009 | ... | 0.001 | ... | ... | 0.09 | $< 10^{-6}$ | |
| Asymm H$\beta$ | −0.14 | 0.09 | 0.21 | −0.03 | 0.12 | 0.07 | −0.22 | −0.32 | −0.19 | −0.03 | 0.13 | −0.36 | −0.50 |
|  | ... | ... | ... | ... | ... | ... | 0.10 | 0.02 | ... | ... | ... | 0.004 | $< 10^{-4}$ |
| EW Fe II | −0.29 | 0.06 | 0.10 | −0.32 | 0.33 | 0.25 | 0.04 | −0.21 | −0.15 | −0.05 | −0.19 | −0.02 | −0.17 |
|  | 0.03 | ... | ... | 0.02 | 0.01 | 0.06 | ... | ... | ... | ... | ... | ... | ... |
| Fe II/H$\beta$ | −0.25 | −0.01 | 0.19 | −0.33 | 0.35 | 0.30 | −0.16 | −0.33 | −0.26 | −0.15 | −0.15 | −0.01 | −0.16 |
|  | 0.06 | ... | ... | 0.01 | 0.008 | 0.03 | ... | 0.01 | 0.05 | ... | ... | ... | ... |
| $\lambda 5007$/H$\beta$ | 0.15 | 0.08 | −0.10 | 0.20 | −0.22 | 0.00 | −0.29 | −0.24 | −0.21 | −0.29 | −0.09 | −0.05 | 0.04 |
|  | ... | ... | ... | ... | 0.10 | ... | 0.03 | 0.07 | ... | 0.03 | ... | ... | ... |
| $M_{OIII}$ | −0.75 | 0.78 | −0.47 | −0.23 | 0.18 | 0.12 | −0.01 | −0.11 | −0.14 | −0.09 | 0.01 | 0.04 | −0.09 |
|  | $< 10^{-6}$ | $< 10^{-6}$ | 0.003 | 0.10 | ... | ... | ... | ... | ... | ... | ... | ... | ... |
| EW $\lambda 5007$ | 0.00 | 0.18 | −0.24 | 0.18 | −0.29 | −0.13 | 0.30 | 0.05 | 0.07 | −0.07 | −0.19 | 0.02 | 0.10 |
|  | ... | ... | ... | ... | 0.03 | ... | 0.02 | ... | ... | ... | ... | ... | ... |
| W50 $\lambda 5007$ | 0.35 | −0.47 | 0.25 | 0.28 | 0.00 | 0.10 | −0.38 | −0.27 | −0.18 | −0.09 | 0.11 | −0.03 | 0.15 |
|  | 0.02 | $< 10^{-3}$ | 0.06 | 0.05 | ... | ... | 0.009 | 0.07 | ... | ... | ... | ... | ... |
| W20 $\lambda 5007$ | 0.19 | −0.40 | 0.44 | 0.03 | 0.14 | 0.15 | −0.18 | −0.27 | −0.21 | −0.08 | 0.06 | 0.03 | −0.18 |
|  | ... | 0.005 | 0.01 | ... | ... | ... | ... | 0.06 | ... | ... | ... | ... | ... |
| W50/W20 | 0.25 | −0.07 | −0.16 | 0.23 | −0.09 | 0.09 | −0.39 | −0.14 | −0.10 | −0.07 | −0.02 | −0.03 | −0.02 |
|  | 0.09 | ... | ... | ... | ... | ... | 0.007 | ... | ... | ... | ... | ... | ... |
| AI20 $\lambda 5007$ | 0.03 | −0.17 | 0.49 | −0.05 | 0.04 | 0.04 | −0.14 | −0.09 | −0.05 | 0.07 | −0.11 | −0.02 | −0.08 |
|  | ... | ... | 0.005 | ... | ... | ... | ... | ... | ... | ... | ... | ... | ... |
| IPV20 $\lambda 5007$ | 0.31 | −0.49 | 0.49 | 0.17 | −0.02 | 0.01 | −0.31 | −0.20 | −0.17 | −0.06 | 0.06 | 0.10 | −0.08 |
|  | 0.03 | $< 10^{-3}$ | 0.005 | ... | ... | ... | 0.04 | ... | ... | ... | ... | ... | ... |
| Asy20 $\lambda 5007$ | −0.17 | 0.09 | 0.31 | −0.07 | 0.01 | 0.03 | 0.02 | 0.04 | 0.09 | 0.12 | −0.23 | −0.06 | −0.02 |
|  | ... | ... | 0.09 | ... | ... | ... | ... | ... | ... | ... | ... | ... | ... |
| $\lambda 5007$/H$\beta_n$ | 0.12 | 0.09 | −0.43 | 0.10 | −0.12 | 0.04 | 0.27 | 0.08 | 0.12 | −0.04 | −0.20 | −0.23 | 0.04 |
|  | ... | ... | 0.01 | ... | ... | ... | 0.05 | ... | ... | ... | ... | 0.10 | ... |
| $L_{core}$ | 0.59 | −0.55 | −0.10 | −0.49 | 0.66 | 0.83 | −0.32 | −0.46 | −0.47 | −0.33 | 0.16 | −0.06 | −0.21 |
|  | $< 10^{-5}$ | $< 10^{-4}$ | ... | $< 10^{-3}$ | $< 10^{-6}$ | $< 10^{-6}$ | 0.02 | $< 10^{-3}$ | $< 10^{-3}$ | 0.02 | ... | ... | ... |
| $L_{extended}$ | 0.58 | −0.48 | 0.09 | 0.32 | −0.48 | −0.01 | −0.19 | −0.03 | 0.02 | −0.02 | 0.21 | −0.06 | 0.00 |
|  | $< 10^{-5}$ | $< 10^{-3}$ | ... | 0.02 | $< 10^{-3}$ | ... | ... | ... | ... | ... | ... | ... | ... |

[a]See § 2 for parameter definitions.
[b]Spearman rank correlation coefficients are tabulated, and with the two-tailed probability of such a correlation arising by chance given beneath (when less than 0.1).

TABLE 6 – *Continued*

|  | Asymm Hβ | EW Fe II | Fe II/Hβ | λ5007/Hβ | $M_{OIII}$ | EW λ5007 | W50 λ5007 | W20 λ5007 | W50/W20 λ5007 | AI20 λ5007 | IPV20 λ5007 | Asy20 λ5007 | $\frac{\lambda 5007}{H\beta_n}$ | $L_{core}$ |
|---|---|---|---|---|---|---|---|---|---|---|---|---|---|---|
| EW Fe II | 0.27 |  |  |  |  |  |  |  |  |  |  |  |  |  |
|  | 0.05 |  |  |  |  |  |  |  |  |  |  |  |  |  |
| Fe II/Hβ | 0.26 | 0.95 |  |  |  |  |  |  |  |  |  |  |  |  |
|  | 0.05 | $< 10^{-6}$ |  |  |  |  |  |  |  |  |  |  |  |  |
| λ5007/Hβ | −0.07 | −0.47 | −0.41 |  |  |  |  |  |  |  |  |  |  |  |
|  | ... | $< 10^{-3}$ | 0.002 |  |  |  |  |  |  |  |  |  |  |  |
| $M_{OIII}$ | 0.18 | 0.34 | 0.32 | −0.38 |  |  |  |  |  |  |  |  |  |  |
|  | ... | 0.01 | 0.02 | 0.003 |  |  |  |  |  |  |  |  |  |  |
| EW λ5007 | −0.22 | −0.41 | −0.47 | 0.77 | −0.38 |  |  |  |  |  |  |  |  |  |
|  | 0.10 | 0.002 | $< 10^{-3}$ | $< 10^{-6}$ | 0.003 |  |  |  |  |  |  |  |  |  |
| W50 λ5007 | 0.03 | 0.19 | 0.25 | −0.03 | −0.30 | −0.15 |  |  |  |  |  |  |  |  |
|  | ... | ... | 0.09 | ... | 0.04 | ... |  |  |  |  |  |  |  |  |
| W20 λ5007 | 0.23 | 0.46 | 0.47 | −0.21 | −0.22 | −0.16 | 0.77 |  |  |  |  |  |  |  |
|  | ... | 0.001 | 0.001 | ... | ... | ... | $< 10^{-6}$ |  |  |  |  |  |  |  |
| W50/W20 | −0.12 | −0.21 | −0.11 | 0.35 | −0.12 | 0.03 | 0.31 | −0.27 |  |  |  |  |  |  |
|  | ... | ... | ... | 0.02 | ... | ... | 0.04 | 0.07 |  |  |  |  |  |  |
| AI20 λ5007 | 0.29 | 0.53 | 0.57 | −0.23 | 0.01 | −0.27 | 0.15 | 0.44 | −0.22 |  |  |  |  |  |
|  | 0.05 | $< 10^{-3}$ | $< 10^{-4}$ | ... | ... | 0.06 | ... | 0.002 | ... |  |  |  |  |  |
| IPV20 λ5007 | 0.09 | 0.39 | 0.42 | −0.20 | −0.25 | −0.26 | 0.66 | 0.79 | −0.15 | 0.45 |  |  |  |  |
|  | ... | 0.007 | 0.004 | ... | 0.09 | 0.08 | $< 10^{-6}$ | $< 10^{-6}$ | ... | 0.001 |  |  |  |  |
| Asy20 λ5007 | 0.33 | 0.47 | 0.45 | −0.09 | 0.09 | −0.07 | −0.07 | 0.20 | −0.14 | 0.73 | −0.04 |  |  |  |
|  | 0.02 | $< 10^{-3}$ | 0.002 | ... | ... | ... | ... | ... | ... | $< 10^{-6}$ | ... |  |  |  |
| λ5007/Hβ$_n$ | −0.31 | −0.17 | −0.26 | 0.12 | −0.08 | 0.31 | −0.17 | −0.38 | 0.04 | −0.25 | −0.34 | −0.16 |  |  |
|  | 0.03 | ... | 0.06 | ... | ... | 0.03 | ... | 0.005 | ... | 0.08 | 0.01 | ... |  |  |
| $L_{core}$ | −0.01 | 0.16 | 0.24 | −0.09 | −0.31 | −0.27 | 0.34 | 0.32 | 0.17 | 0.04 | 0.23 | −0.08 | −0.08 |  |
|  | ... | ... | 0.09 | ... | 0.03 | 0.05 | 0.02 | 0.03 | ... | ... | ... | ... | ... |  |
| $L_{extended}$ | −0.08 | −0.12 | −0.08 | 0.11 | −0.44 | −0.02 | 0.39 | 0.16 | 0.32 | −0.01 | 0.29 | −0.14 | 0.09 | 0.27 |
|  | ... | ... | ... | ... | 0.001 | ... | 0.006 | ... | 0.03 | ... | 0.05 | ... | ... | 0.05 |

TABLE 7

Projections of PCA Eigenvectors onto Line and Continuum Properties[a]

| Property[b] | Vector 1 27.4% | Vector 2 43.8% | Vector 3 58.4% | Vector 4 71.3% | Vector 5 81.5% | Vector 6 88.0% |
|---|---|---|---|---|---|---|
| Log $R_V$ | −0.443 | 0.135 | 0.156 | −0.018 | 0.621 | 0.223 |
| Mabs | 0.048 | 0.137 | −0.223 | −0.752 | 0.225 | −0.539 |
| EW H$\beta$ | 0.406 | −0.239 | 0.087 | −0.348 | 0.346 | 0.412 |
| FWHM H$\beta$ | 0.452 | −0.417 | −0.199 | −0.010 | −0.080 | 0.046 |
| Shift H$\beta$ | −0.178 | −0.382 | −0.392 | 0.402 | 0.371 | −0.460 |
| Shape2 H$\beta$ | 0.382 | 0.297 | 0.377 | 0.205 | −0.069 | −0.463 |
| Asymm H$\beta$ | −0.370 | −0.236 | −0.300 | −0.244 | −0.496 | 0.072 |
| $\lambda 5007$/H$\beta$ | 0.007 | 0.637 | −0.441 | 0.016 | −0.091 | 0.171 |
| Fe II/H$\beta$ | −0.343 | −0.190 | 0.546 | −0.223 | −0.201 | −0.166 |

[a]This non-parametric analysis includes 53 objects.
[b]See § 2.3.1 for parameter definitions.

TABLE 8
PROJECTIONS OF PCA EIGENVECTORS ONTO LINE AND CONTINUUM PROPERTIES[a]

| Property[b] | Vector 1 21.3% | Vector 2 40.1% | Vector 3 54.2% | Vector 4 67.6% | Vector 5 78.0% | Vector 6 83.4% |
|---|---|---|---|---|---|---|
| $\alpha_{ox}$ | $-0.439$ | $-0.247$ | $-0.002$ | 0.238 | 0.178 | 0.305 |
| Log $R_V$ | 0.045 | 0.405 | 0.139 | $-0.226$ | $-0.442$ | 0.217 |
| Mabs | 0.344 | 0.105 | $-0.146$ | $-0.453$ | 0.185 | $-0.164$ |
| EW H$\beta$ | 0.328 | $-0.394$ | 0.104 | $-0.191$ | $-0.152$ | 0.022 |
| FWHM H$\beta$ | 0.101 | $-0.483$ | $-0.302$ | 0.165 | 0.103 | $-0.294$ |
| Shift H$\beta$ | $-0.113$ | $-0.030$ | $-0.465$ | 0.219 | $-0.429$ | 0.146 |
| Shape2 H$\beta$ | 0.127 | $-0.209$ | 0.480 | 0.194 | 0.241 | 0.077 |
| Asymm H$\beta$ | $-0.271$ | 0.054 | $-0.491$ | $-0.300$ | 0.208 | 0.135 |
| $\lambda 5007$/H$\beta$ | 0.082 | 0.401 | 0.027 | 0.108 | 0.546 | 0.231 |
| Fe II/H$\beta$ | $-0.382$ | $-0.046$ | 0.240 | $-0.324$ | $-0.082$ | $-0.541$ |
| W50/W20 $\lambda 5007$ | 0.062 | 0.350 | $-0.215$ | 0.353 | 0.155 | $-0.567$ |
| IPV20 $\lambda 5007$ | $-0.476$ | 0.136 | 0.250 | 0.101 | $-0.099$ | $-0.165$ |
| Asy20 $\lambda 5007$ | $-0.290$ | $-0.156$ | $-0.050$ | $-0.443$ | 0.283 | 0.079 |

[a]This non-parametric analysis includes 30 objects.
[b]See § 2.3.1 and § 2.3.2 for parameter definitions.

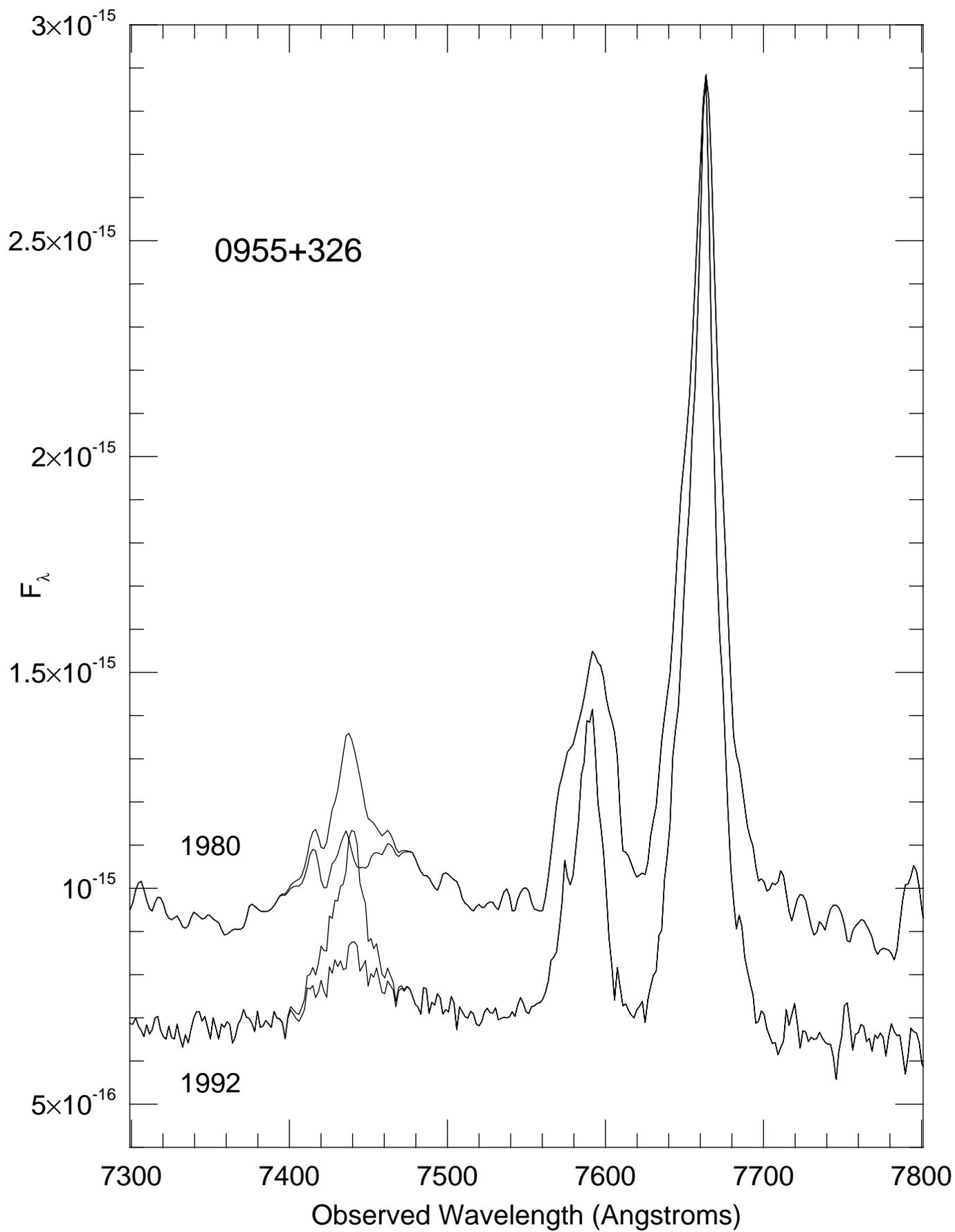

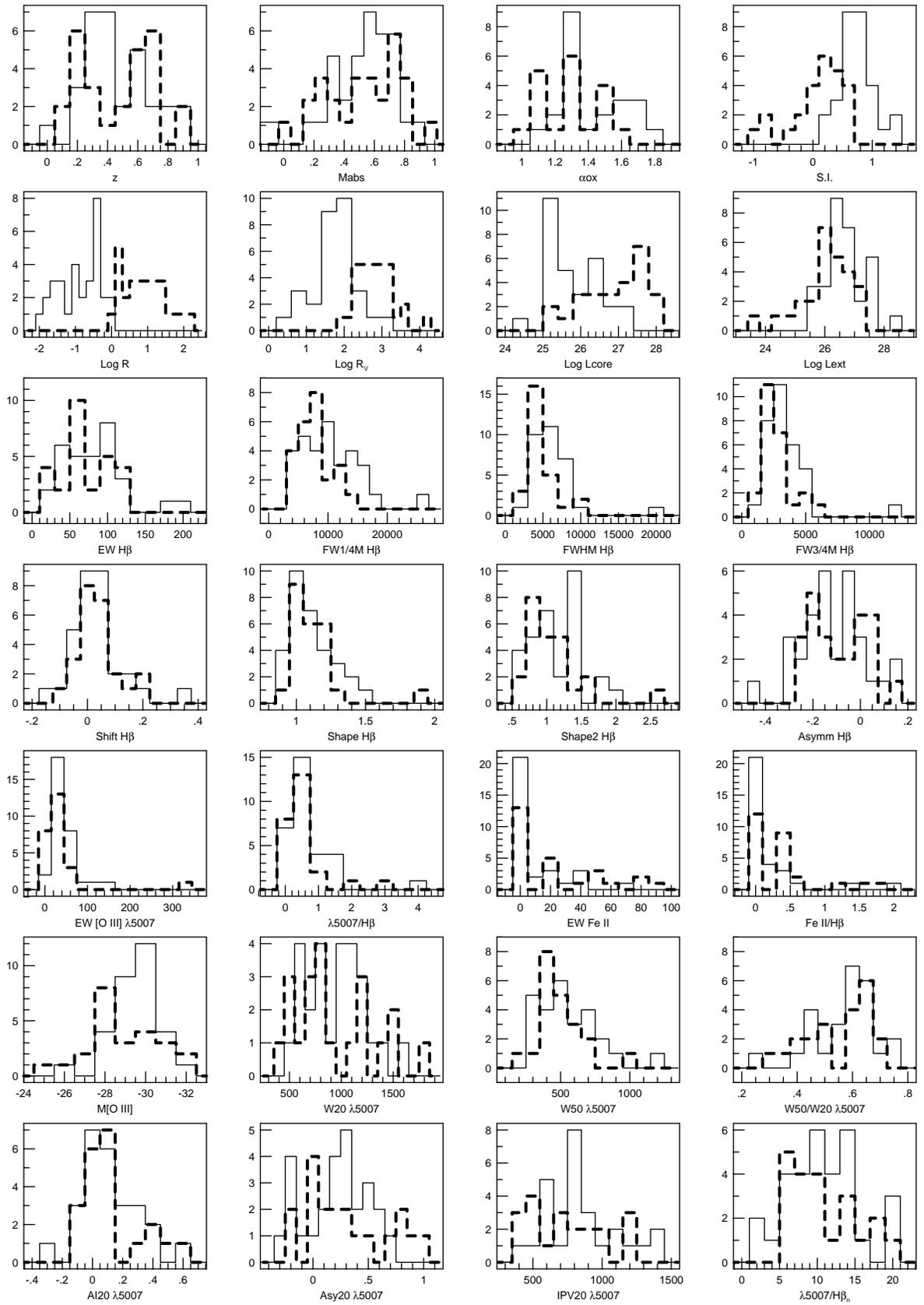

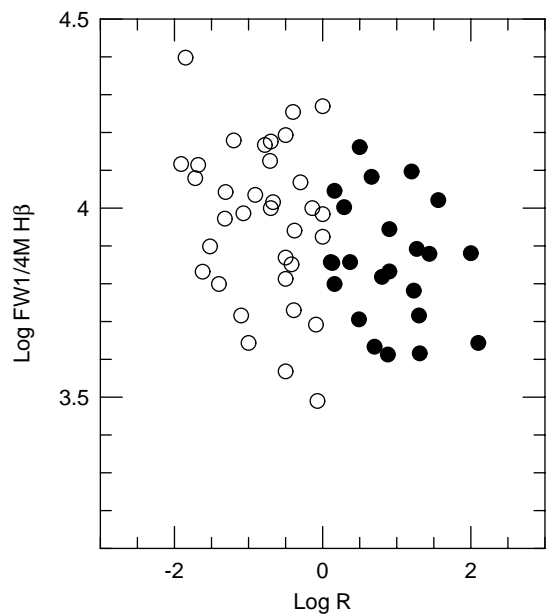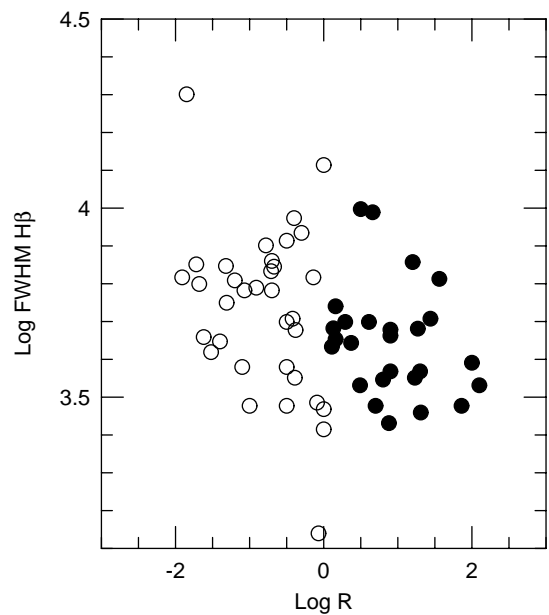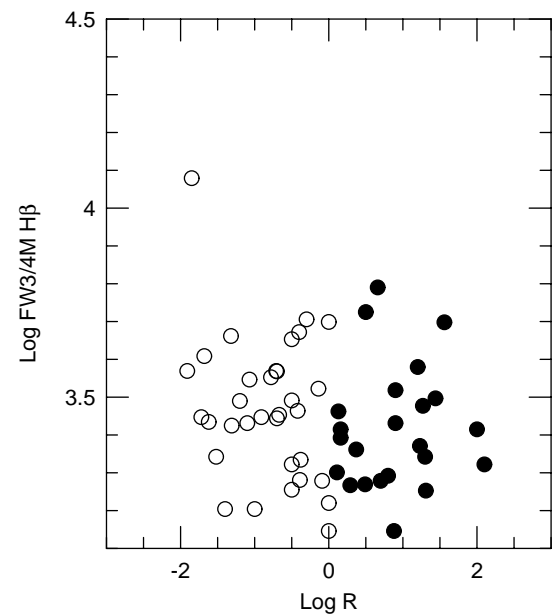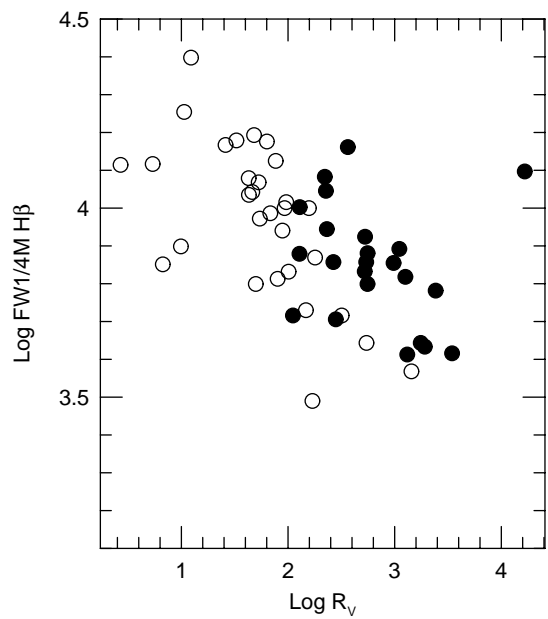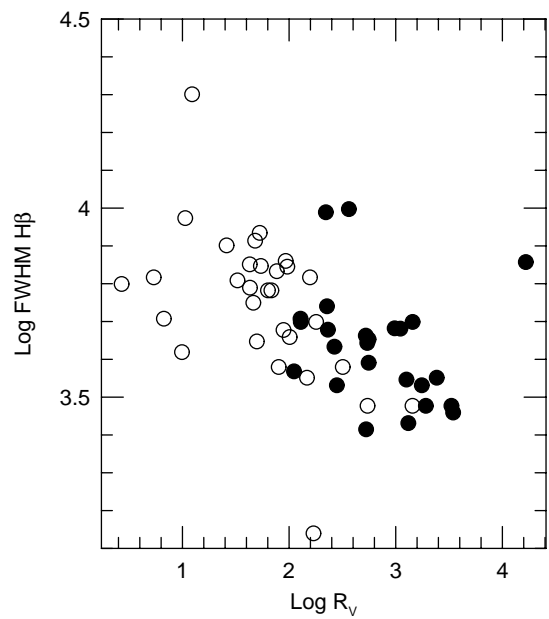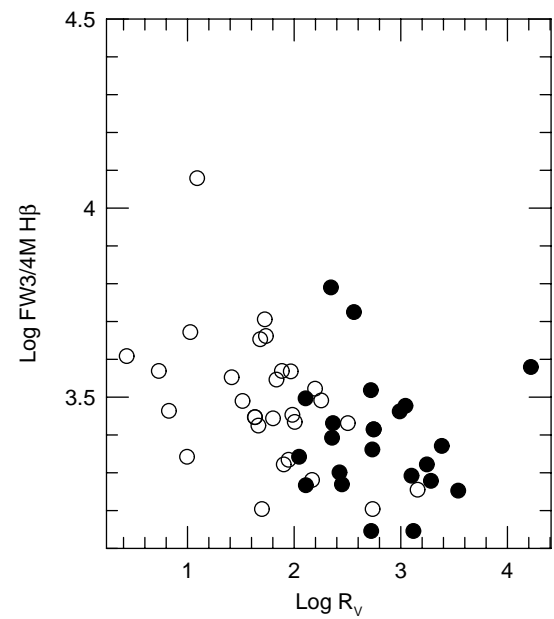

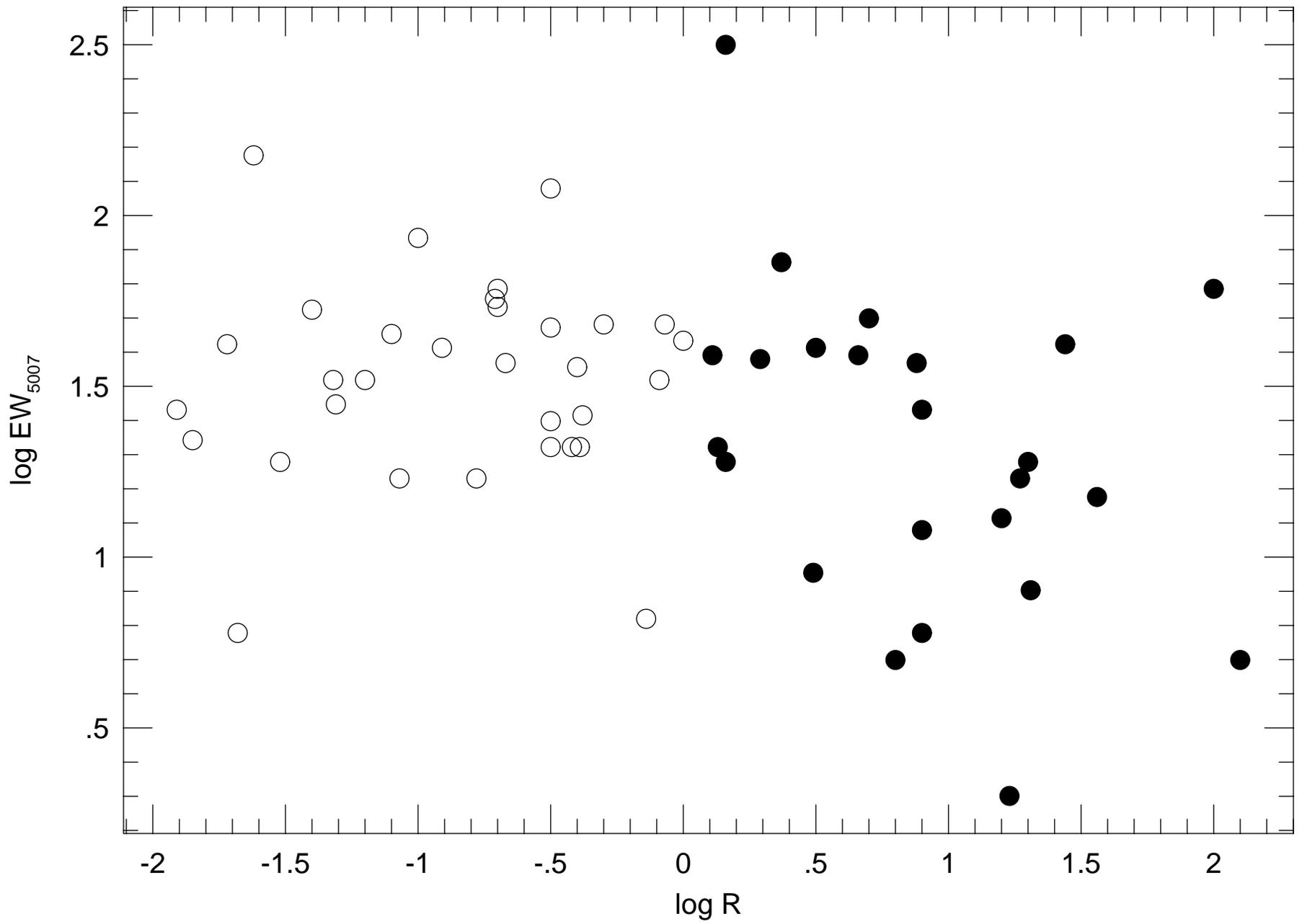

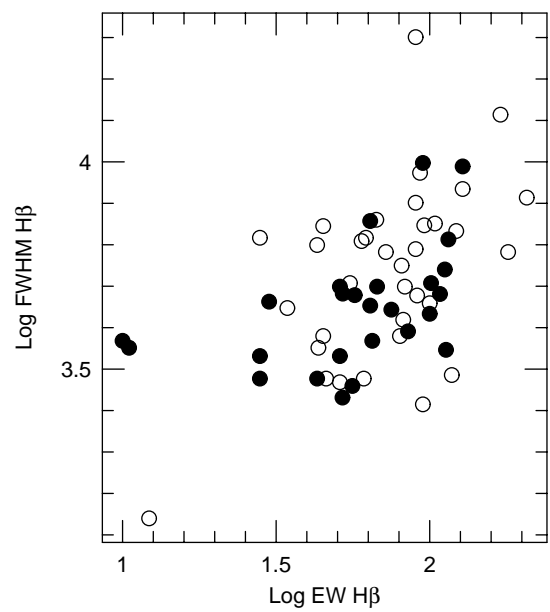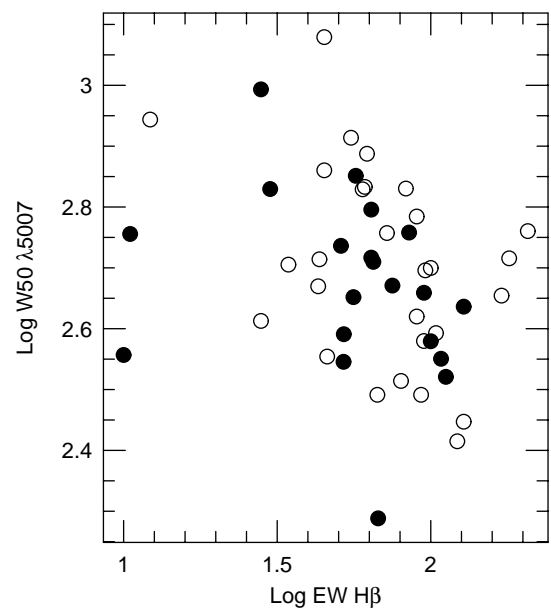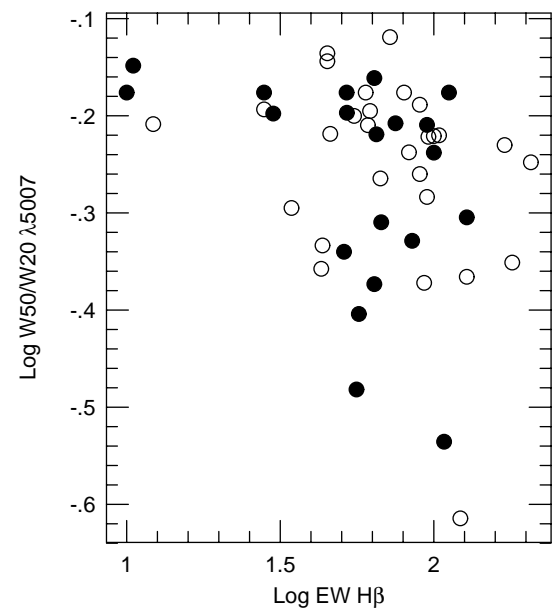

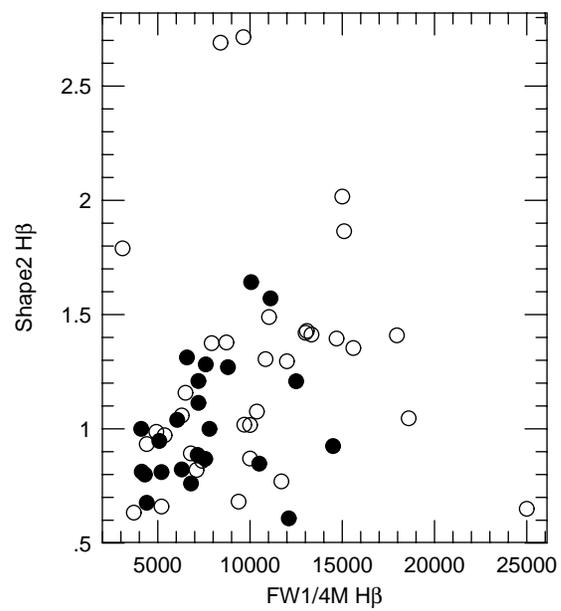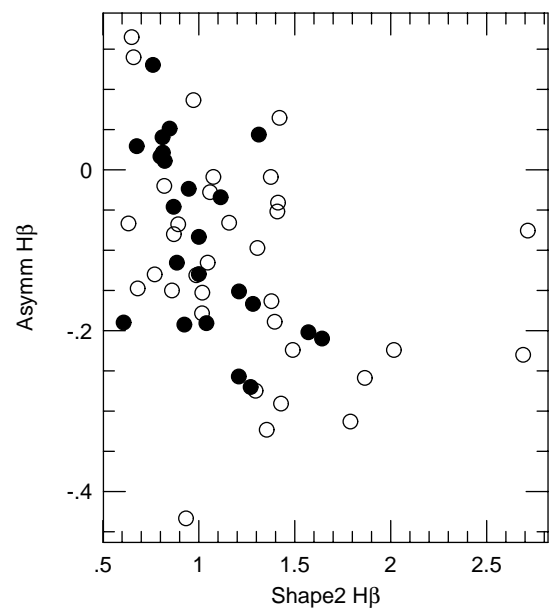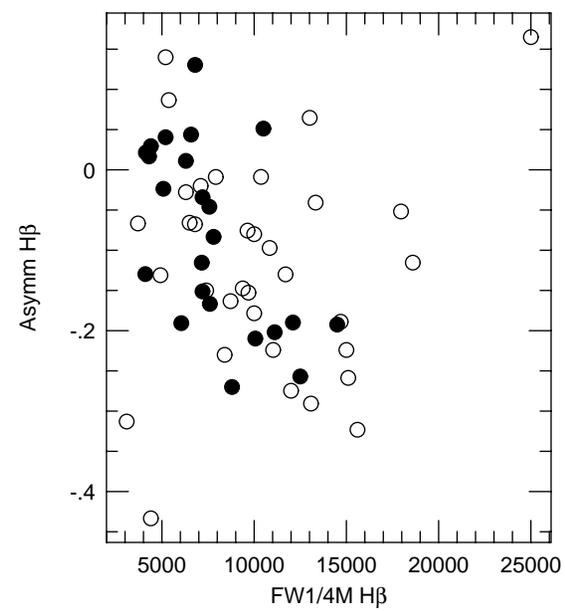

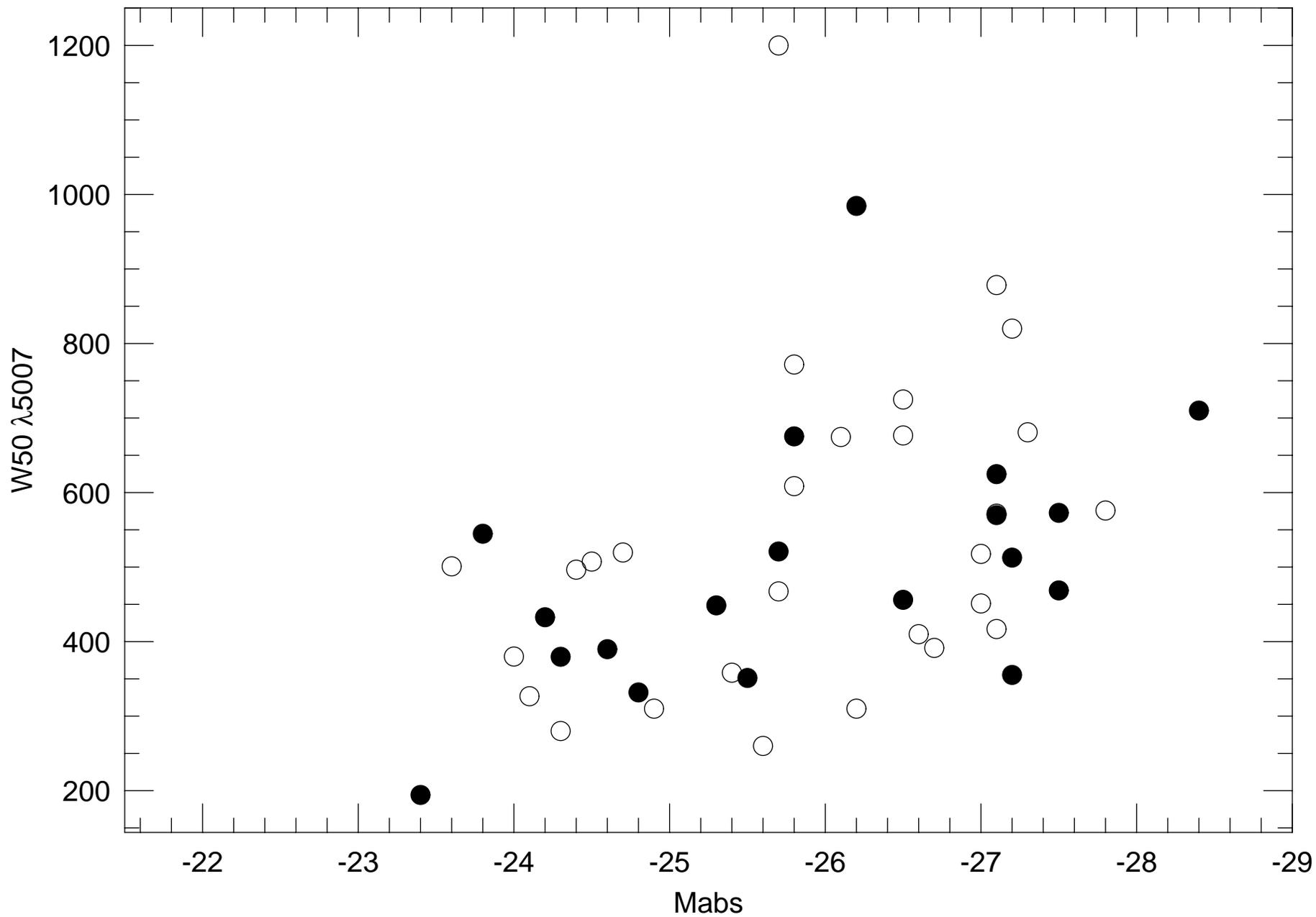

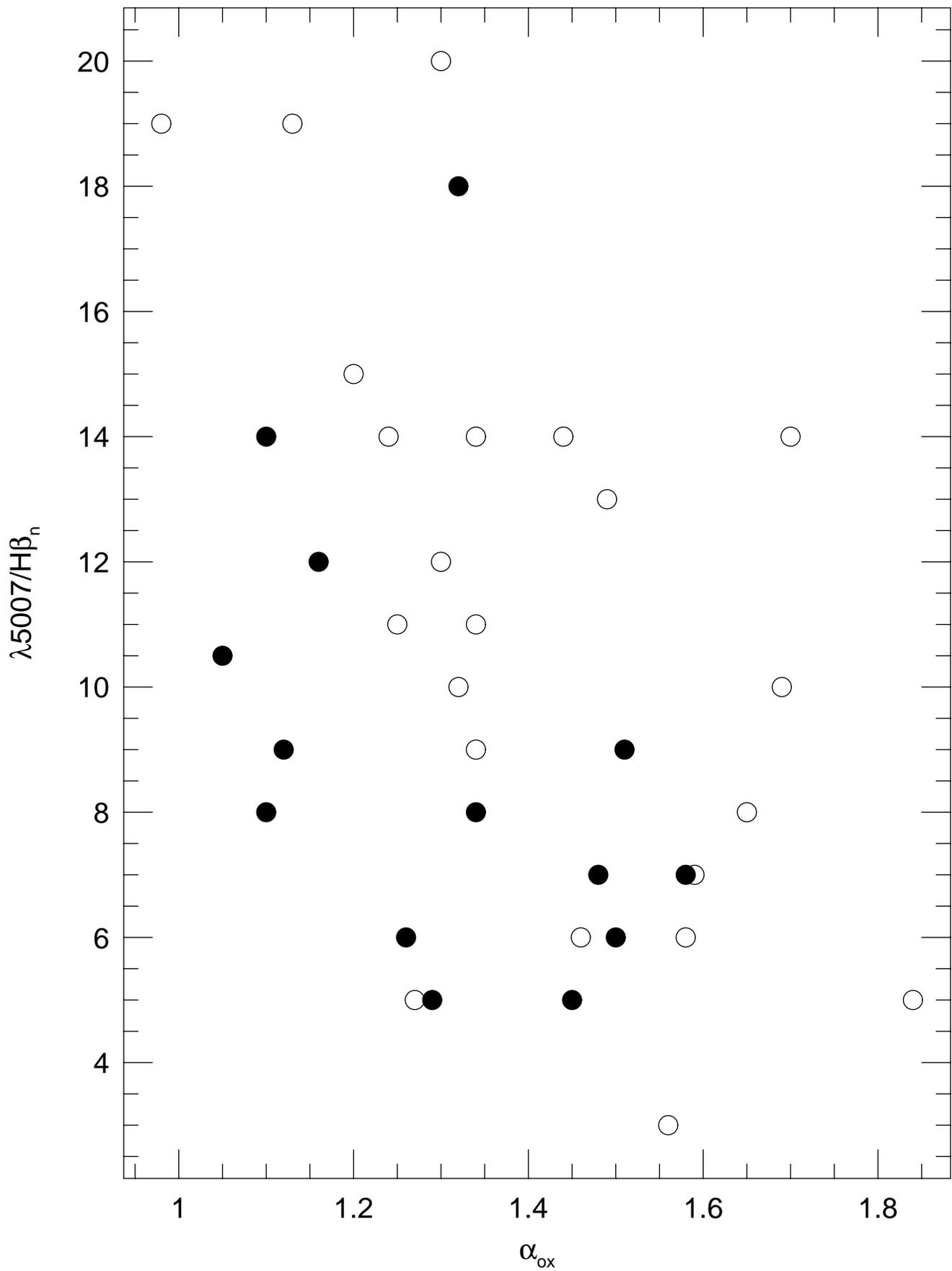

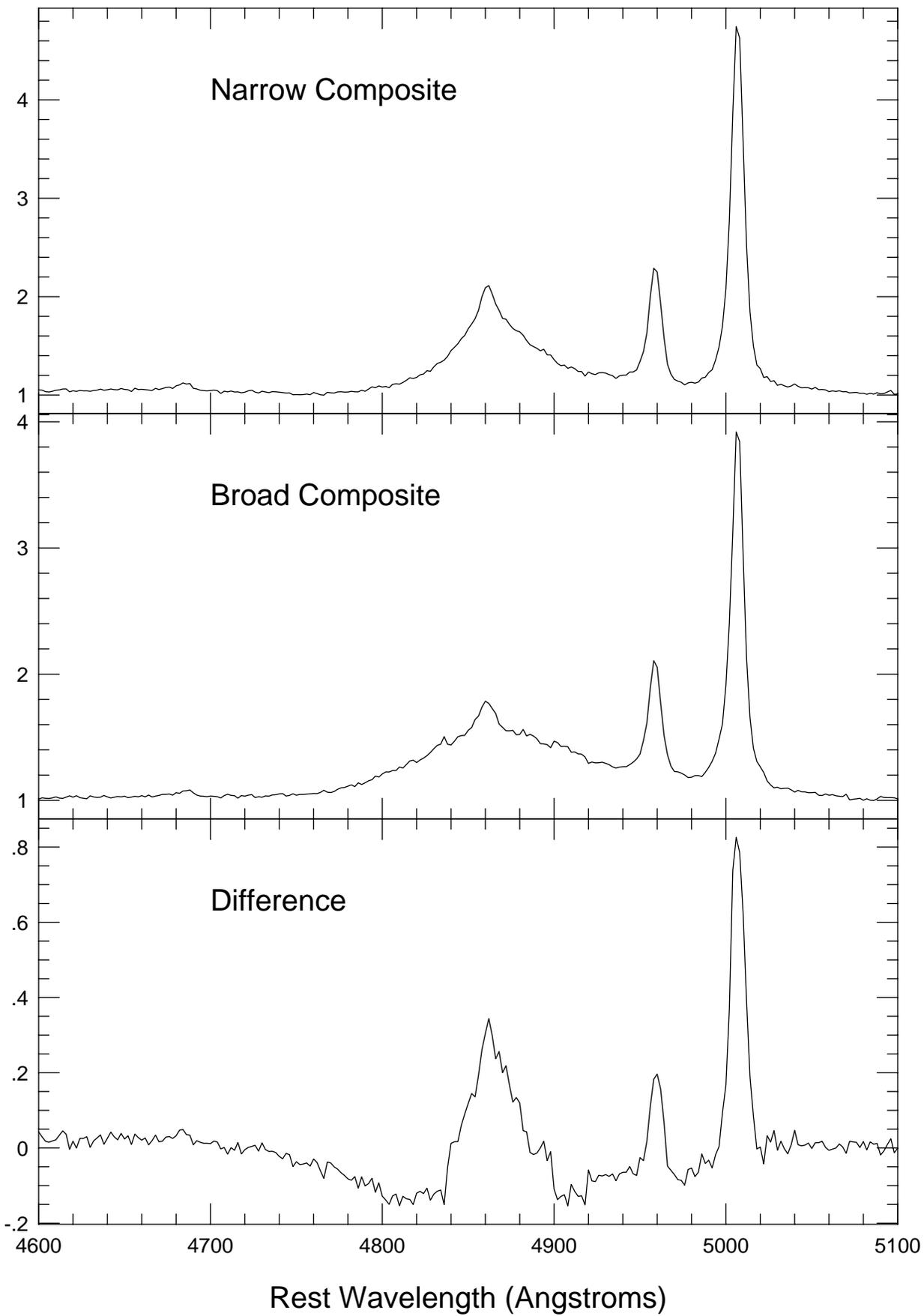